\documentclass[%
 amsmath,amssymb,
 aps,
 prb,
 twocolumn,
 superscriptaddress,
 showpacs,
 showkeys,
]{revtex4-1}

\usepackage{hyperref}
\usepackage{graphicx}
\usepackage{dcolumn}
\usepackage{bm}
\usepackage{xcolor}
\usepackage{color}
\usepackage{mathrsfs}
\usepackage{braket}
\usepackage{float}
\begin{document}

\title{A scalar photon theory for near-field radiative heat transfer} 

\author{Jiebin Peng}
\affiliation{Department of Physics, National University of Singapore, Singapore 117551, Republic of Singapore}
\author{Han Hoe Yap}
\affiliation{NUS Graduate School for Integrative Sciences and Engineering, Singapore 117456, Republic of Singapore}
\author{Gang Zhang}
\affiliation{Institute of High Performance Computing, A*STAR, Singapore 138632, Republic of Singapore}
\author{Jian-Sheng Wang}
\affiliation{Department of Physics, National University of Singapore, Singapore 117551, Republic of Singapore}

\date{\today}

\begin{abstract}
We study a one-dimensional model of radiative heat transfer for which the effect of the electromagnetic field is only from the scalar
potential and thereby ignoring the vector potential contribution.  This is a valid assumption when the distances between objects
are of the order of nanometers.  Using Lorenz gauge, the scalar field is quantized with the canonical 
quantization scheme, {giving} rise to scalar photons.  In the limit as the speed of light approaches infinity,
the theory reduces to a pure Coulomb interaction governed by the Poisson equation. 
The model describes very well parallel plate capacitor physics, where a new length scale related to its 
capacitance emerges. Shorter than this length scale we see greater radiative heat transfer. This differs markedly from 
the usual Rytov fluctuational electrodynamics theory in which the enhancement is due to evanescent modes shorter than the thermal wavelengths. Our theory may explain recent experiments where charge fluctuations
instead of current fluctuations play a dominant role {in radiative heat transfer}. Finally, due to the asymmetric electron-bath couplings, thermal rectification effects are also observed and reported.
\end{abstract}
\pacs{05.60.Gg, 44.40.+a, 12.20.-m}
\keywords{radiative heat transfer, scalar photons}
\maketitle

\section{Introduction}
Planck's theory of blackbody radiation\citep{Planck_book} lays the foundation for radiative heat transfer\cite{howellbook}.  Planck himself was aware that his theory does not apply when distances between objects are comparable to the thermal
wavelength, which is of the order of micrometers at room temperature.  However, it was not until the early 1970s that scientists started to look into this problem of near-field radiation qualitatively. Rytov\cite{Rytov} developed a general theory in the 1950s, known as
fluctuational electrodynamics. However, Rytov did not look at near-field radiative heat transfer closely.  It was
Polder and von Hove (PvH) who gave the first formula for the heat transfer in a parallel-plate geometry\cite{PvH,hargreaves1973radiative}, which assumes a Landauer form in the parlance of mesoscopic transport theory\cite{yap16}. 
Also, the electromagnetic
field is treated classically. The quantum effect is put in only at the last step where the quantum version of the 
fluctuation-dissipation theorem is evoked. This theory has been applied later to study the heat transfer in various other geometries, such as between a sphere and a substrate\cite{sphere}, between a cylinder and a perforated surface\cite{cylinder}, as well as between one-dimensional gratings\cite{gratings}. Under the context of photonic thermal management, the near-field radiative heat transfer (NFRHT)\cite{boriskina2016heat,hsu2016entropic} between different 2D materials are studied\cite{PhysRevB.85.155422,rodriguez2015radiative,peng2015thermal}, while novel concepts are frequently proposed, e.g. vacuum thermal rectifier\cite{otey2010thermal}, near-field thermal transistor\cite{PhysRevLett.112.044301} and radiative thermal memory device\cite{PhysRevLett.113.074301}.  Most recently, precision measurements of radiative heat transfer (RHT) in nanoscale gaps were achieved in different materials with plane-plane or tip-plane geometry\cite{Shen09,Kloppstech2015,kim2015radiative,st2016near,song2016radiative}. In brief, sixty years after its birth, fluctuational electrodynamics continues to stimulate interests across a broad spectrum of the scientific community.

With the exception of a few works\cite{mahantunneling,greffet_coulomb,yu,xiong14}, the quantity of interest has always been the electromagnetic energy flux density, commonly known as the Poynting vector\cite{PhysRevB.85.155422,rodriguez2015radiative,peng2015thermal}. Further, one frequently considers current or polarization density to be the only fluctuating source responsible for heat transfer\cite{sphere,cylinder,gratings,joulain2005surface,volokitin07}. While such an approach, namely, fluctuational electrodynamics with current as source, has been very successful, some questions persist: down to which length scale does a semi-classical theory as such remain valid? Is current fluctuation the only mechanism to be accounted for in electromagnetic heat transfer? Such are the issues we wish to address in the present work. Based on a double electron dot model, we give a detailed account of the fully quantum-mechanical treatment of thermal radiation proposed in a recently submitted paper\cite{wang2016microscopic}. 
We focus on the scalar field, which was initially thought to be for the sake of simplicity. However, ignoring the vector potential $\bm{A}$ (which arises from current fluctuation) reveals that charge fluctuation (to which is associated the scalar potential) plays an equally if not more important role in the RHT within short distances. In particular, we identify a length scale much smaller than the thermal wavelength at room temperature. 

From a technical viewpoint, the problem of heat transfer has to be treated as an open system, since in steady 
state, we need a source capable of supplying energy for an indefinite amount of time. On the other hand, for two bodies placed extremely close to each other ($<$10 nm), which is now experimentally feasible\cite{Shen09,Kloppstech2015,kim2015radiative,st2016near,song2016radiative}, a fully-quantum description is needed. In this regard, nonequilibrium Green's function (NEGF) is the natural choice of method. It has been used to study quantum thermal transport of electrons and phonons {\cite{lu2007,jswpre07,wang08review,zhanglifa2013,wang14rev}}, and we wish to extend this method to the case of photon-mediated thermal transport. Importantly, we go beyond the ballistic treatment and consider the nonlinear interactions between the field and the electrons.

This paper is structured as follows. We begin by constructing a two-dot capacitor model in Sec.~\ref{sec:THEORY}, where we also write down the Hamiltonian and discuss the quantization of electrodynamics. In Sec.~\ref{sec:NEGF_Dyson}, we outline the NEGF methods, solve the Dyson equation and discuss the Keldysh equation. In Sec.~\ref{sec:Currents}, we quantize a ``Poynting scalar'' (heat flux due to the Coulomb interaction) and relate its expectation value to a Green's function. In Sec.~\ref{sec:Currents_var_limit}, we analyze the current expression under different limiting procedures to draw physical insights from our simple model. In Sec.~\ref{sec:Self_energy}, we tackle the problem of self energy calculations. Three approximations are discussed, putting into perspective Rytov's theory and our NEGF approach. In Sec.~\ref{RESULT}, numerical simulations of our theory are given, allowing us to investigate two-dot transport behaviors including thermal rectification. We conclude and summarize in Sec.~\ref{sec:Conclusion}.


\section{The model and quantization} \label{sec:THEORY} 

In this paper, we deal exclusively with a one-dimensional field.  By one dimension, we do not mean that our physical system is a one-dimensional line.  Instead, we assume the fields (the scalar potential, or the electric field) which live in a three-dimensional space, depend only on one single variable $z$.  Thus, a three-dimensional parallel plate with sufficiently large
cross-sectional area $A$ belongs to a one-dimensional problem. 

\begin{figure}[htp] 
  \centering
  \includegraphics[totalheight=60mm]{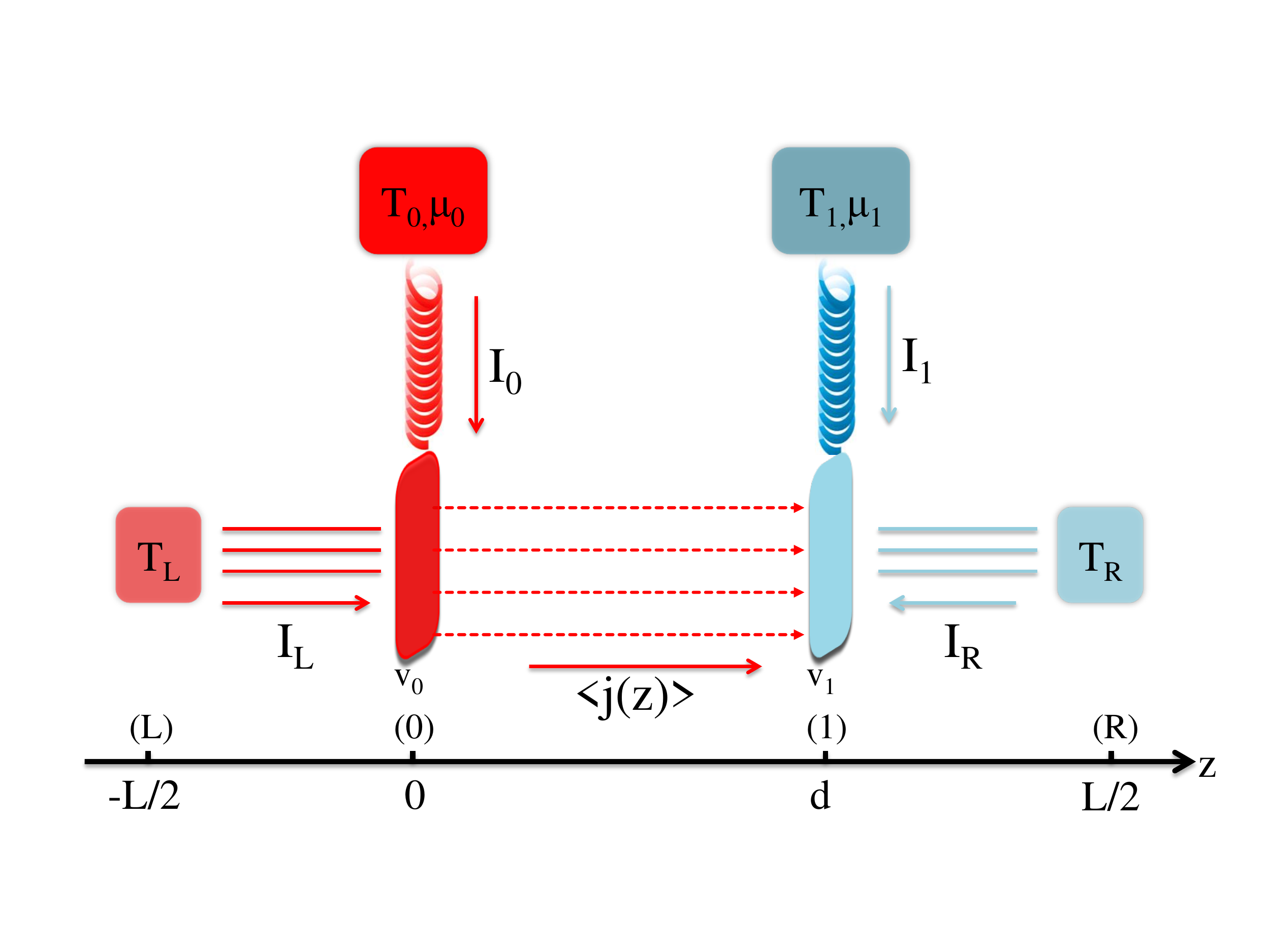}
  \caption{Schematic of the two-dot capacitor model. Scalar photons permeate the space, with $T_L$ and $T_R$ indicating the photon bath temperatures. Electrons of onsite energies $v_0$ and $v_1$ are located at $z_0=0$ and $z_1=d$, connected to fermionic reservoirs at temperatures and chemical potentials $(T_0,\mu_0)$ and $(T_1,\mu_1)$ respectively.}
 \label{Two_dots}
\end{figure} 

\subsection{The model}
We consider two quantum dots, which can be thought of as nanoscale parallel plates with a possible charge of either
0 or $-Q$ as shown in Fig.~\ref{Two_dots}. To each electron is connected a reservoir, allowing its charge to fluctuate and generate radiation. This simple model allows for analytically tractable expressions, all the while providing an essential ingredient for radiative heat transfer: two separated systems connected to two baths. We postulate the Lagrangian as  $\mathcal{L} = L_e + L_\gamma + L_{\mathrm{int}}$, with
\begin{align}
\begin{split}
L_e =& \sum_{j=0,1} c_j^\dagger \bigl(i\hbar  \dot{c}_j - v_j c_j\bigr)  \\
&+\sum_{j=0,1}\sum_{k\in\textrm{bath}} d_{jk}^\dagger(i\hbar \dot{d}_{jk} -\epsilon_{jk} d_{jk}) \\
& -\sum_{j=0,1}\sum_{k\in\textrm{bath}} \left(V_{jk}c_j^\dagger d_{jk} + \textrm{H.c.}\right), \\
L_\gamma =& s \int dz \left[  -\frac{1}{2}\dot{\phi}^2 + \frac{c^2}{2} \left( \frac{\partial \phi}{\partial z} \right)^2 \right],\\
L_{\mathrm{int}} =& -  \sum_{j=0,1} (-Q) c_j^\dagger c_j \phi(z_j).
\end{split}
\end{align} 
Here we assume a tight-binding model for the electrons represented by fermionic annihilation operators $c_j$ and creation
operators $c_j^\dagger$.  The electrons are located at positions $z_j$ with $z_0 = 0$ and
$z_1 = d$. The onsite energy of dot $j$ is $v_j$. The electrons themselves at different sites do not have direct hopping coupling but the electrons are coupled to their
respective baths. Electron bath $j$ is described by fermionic operators $d^\dagger_{jk}$ with energy $\epsilon_{jk}$ and is coupled to dot $j$ via a tunneling amplitude $V_{jk}$, where $k$'s are the reservoir modes. For the field $\phi(z,t)$,  we have defined the scale factor $s=\epsilon_0 A/c^2$, where $\epsilon_0$ is
the vacuum permittivity, $A$ is the cross-sectional area of the plates, $c$ is the speed of light, and the
integral extends from $-\infty$ to $\infty$.   We split this integral into three parts, $(-\infty, -L/2]$, $[-L/2,L/2]$, and $[L/2, +\infty)$,
and consider them to be the left photon bath, central region, and right photon bath. 

The most striking feature of the Lagrangian is the scalar field part.  Since we can split the Lagrangian
as $\mathcal{L}=T-V$, the kinetic energy minus the potential energy, we see that the Lagrangian for the field has both a negative
kinetic energy and a negative potential energy.   This is the correct Lagrangian to use since it gives the wave equation with the charge as the source from the principle of least action ($\delta \int \mathcal{L} dt = 0$)\cite{tannoudji1989photons,keller2011quantum},
\begin{equation}
\frac{1}{c^2}\ddot{\phi} -  \frac{\partial^2 \phi}{\partial z^2} = \frac{\rho}{\epsilon_0} =  \frac{1}{\epsilon_0 A} \sum_{j} (-Q) c_j^\dagger c_j \delta(z-z_j).
\end{equation}
We note that in the limit $c\to \infty$, it reduces to the (one-dimensional) Poisson equation for the 
potential.

\subsection{Canonical quantization and Hamiltonian}\label{justification}
The introduction of the Lagrangian gives us a good starting point to quantize the system according to the canonical quantization 
scheme.  We compute the conjugate momenta for the dynamical variables $c_j$, $c_j^\dagger$, and $\phi$:
\begin{align}
\begin{split}
P_{c_j} &= { \partial \mathcal{L} \over \partial \dot{c}_j} =  i\hbar c_j^\dagger, \\
P_{c_j^\dagger} &= { \partial \mathcal{L} \over \partial \dot{c}_j^\dagger} = 0, \\
\Pi_\phi(z) &= { \delta \mathcal{L} \over \delta \dot{\phi}(z) } = -s \dot{\phi}(z). 
\end{split}
\end{align} 
The last derivative above is a functional derivative since $\dot{\phi}(z)$ is a field that depends continuously on $z$.

We impose canonical commutation relations to quantize the system, turning numbers into operators.  However, the fermionic degrees of freedom are already in a quantized form. More precisely, we should think of $c_j$ and its Hermitian conjugate as Grassmann numbers obeying the anti-commutation relation, $c_j c_k^\dagger + c_k^\dagger c_j = \delta_{jk}$. For the field, we have: 
\begin{align}
\begin{split}
\label{eq-comm-phi}
[ \phi(z), \phi(z') ]  &= 0, \\
[ \Pi_\phi(z), \Pi_\phi (z') ] &= 0, \\
[ \phi(z), \Pi_\phi(z') ] &= i\hbar \delta(z-z'). 
\end{split}
\end{align} 
Due to the negative-definite kinetic energy term, the last commutation relation, i.e. $[ \dot{\phi}(z), \phi(z') ] = (i\hbar/s) \delta(z-z')$, differs from the usual ones for phonons (or transverse photons) by
a minus sign. 


The quantum Hamiltonian is obtained from the Legendre transform $H = \sum_j \bigl( P_{c_j} \dot{c}_j + 
P_{c_j^\dagger} \dot{c}_j^\dagger  \bigr) + \int dz\, \Pi_\phi(z) \dot{\phi}(z)  - \mathcal{L}$, giving  $H=H_\gamma + H_e + H_{\mathrm{int}}$\cite{wang2016microscopic}, with:
\begin{align}
\begin{split}
 H_\gamma =& -s \int dz \frac{1}{2} \left[ \dot{\phi}^2+c^2\left(\frac{\partial \phi}{\partial z}\right)^2 \right],\\
 H_e=& \sum_{j=0,1} v_jc_j^\dagger c_j + \sum_{j=0,1}\sum_{k\in\textrm{bath}}\epsilon_{jk} d^\dagger_{jk} d_{jk}  \\
 &+\sum_{j=0,1}\sum_{k\in\textrm{bath}}\left( V_{jk} c_j^\dagger d_{jk} + \textrm{H.c.} \right),\\
H_{\mathrm{int}} =& \sum_{j} (-Q) c^\dagger_j c_j \phi(z_j),
\end{split}\label{eq:Hamiltonian}
\end{align} 
where $H_\gamma$ is the free scalar photon Hamiltonian. In our model, we regard the photon field as a scalar wave propagating at the speed of light, and restrict the 3D photon field to an infinite cuboid with cross-sectional area $A$. From the free photon Hamiltonian, we see that the scalar field obeys a wave equation: $\frac{\partial^2 \phi}{\partial z^2} - \frac{1}{c^2} \frac{\partial^2 \phi}{\partial t^2}=0$ in free space.  $H_{\mathrm{int}}$ is the interaction between electrons and the scalar potential which assumes the form $q_j \phi(z_j)$, where the charge operator at site $j$ is given by $q_j=(-Q)c_j^\dagger c_j$. 

We discuss here the validity of our models. The first question is the use of Lorenz gauge without vector potential---this seems to be incompatible with the gauge condition, $\dot{\phi}/c^2 + \bm{\nabla}\cdot\bm{A} = 0$. 
However, recall that we wish to focus on charge fluctuation, so we shall forego completely the transverse\footnote{For a smooth vector field $\bm{V}$ admitting Helmholtz decomposition\cite{keller2011quantum}, i.e. $\bm{V}=\bm{V}_\perp+\bm{V}_\parallel$, its transverse and longitudinal part satisfy respectively $\bm{\nabla}\cdot\bm{V}_\perp=0$ and $\bm{\nabla}\times\bm{V}_\parallel=\bm{0}$.} current $\bm{J}_\perp=\bm{0}$\footnote{The longitudinal current $\bm{J}_\parallel$ is needed for charge continuity equation but does not appear directly in the Hamiltonian.}. This way, the only part left of the vector potential $\bm{A}=\bm{A}_\perp+\bm{A}_\parallel$ is its longitudinal part $\bm{A}_\parallel$. Now, in the limit $c \to \infty$, one has $\bm{\nabla}\cdot\bm{A}_\parallel=0$, which together with $\bm{A}_\perp=\bm{0}$ implies a vanishing $\bm{A}$. Thus it is possible to work under Lorenz gauge with a zero vector potential, provided that the current is irrotational, $\bm{\nabla}\times\bm{J}=\bm{0}$, and that we take $c\rightarrow \infty$ at the end. Yet a finite speed $c$ is needed for canonical quantization, for otherwise there will be no generalized velocity $\dot{\phi}$ in the Hamiltonian which will result in a vanishing conjugate momentum $\Pi_\phi=0$.

The second question is how to deal with a negative-definite Hamiltonian $H_\gamma$? One does not encounter this problem in the usual quantum electrodynamics\cite{weinberg}, because the negative-definite scalar photon Hamiltonian gets canceled exactly by its longitudinal counterpart that arises from the vector potential\cite{tannoudji1989photons,keller2011quantum}, resulting in a total free photon Hamiltonian that remains positive definite. Here, we circumvent this difficulty by assigning negative temperatures to the photon baths. This will be discussed in greater detail in Appendix \ref{anti_commute}. However, in the limit $c\rightarrow \infty$, the Coulomb interaction---which does not propagate ---is recovered, and the photon baths being placed at $\pm\infty$ are immaterial in actual calculations.

\section{NEGF and Dyson equation}\label{sec:NEGF_Dyson}

In Rytov's theory\citep{Rytov}, one focuses on the electromagnetic fields due to fluctuating sources whose autocorrelation function is given phenomenologically. On the other hand, NEGF focuses on correlation functions and allows fields, matter and their interaction to be studied altogether. As discussed below, the two approaches are equivalent
under the local equilibrium approximation. However, NEGF can handle electron-photon interactions in a perturbative way or through mean-field
schemes such as the self-consistent Born approximation.  Thus, NEGF is a more general and more powerful method.

In the formalism of NEGF, we define the contour-ordered Green's function for photons\cite{wang08review,wang14rev,haug96}:
\begin{align}
\begin{split}
 D(z,\tau;z',\tau') & = \frac{1}{i\hbar} \bigl\langle T_\tau \phi^H(z,\tau) \phi^H(z',\tau') \bigr\rangle_{H} \\
 &\!\!\!\!\!\!\!\!\!\!\!\!\!\!\!\!\!\!\!\!\!\!\!\! =\frac{1}{i\hbar} \bigl\langle T_\tau \phi(z,\tau) \phi(z',\tau') e^{-\frac{i}{\hbar}\!\int\! H_{\mathrm{int}}(\tau'') d\tau'' }\! \bigr\rangle_{H_\gamma+H_e},
 \end{split}\label{photonD}
 \end{align}
 where the first line is in the Heisenberg picture with time evolution according to the total Hamiltonian $H$, while
 on the second line, we have transformed the variables into the  interaction picture.  The interaction Hamiltonian is
\begin{equation}
H_{\mathrm{int}}(\tau)=\sum_j (-Q) c^\dagger_j(\tilde{\tau})c_j(\tau)\phi(z_j,\tau).
\end{equation}
Above, $\tau$ and $\tau'$ are Keldysh contour times. $\tilde{\tau} = \tau + \Delta \tau$. Under $T_\tau$,
i.e., the contour-ordering operator, $\tilde{\tau}$ is always slightly later than $\tau$.
This is to avoid swapping the number operator $c_j^\dagger c_j$ to
$c_j c_j^\dagger$.
$\langle  \dots \rangle = {\rm Tr}(\rho_C \rho_B \rho_\gamma \dots)$ is the product initial state (center system, electron baths, and free 
photon system) and the thermal state is assumed to be of the Gibbs form  $\propto e^{-\beta_i H_i}$.  Particularly noteworthy
is that we have already incorporated the effect of the scalar photon baths in the distribution $\rho_\gamma$, as well
as the effect of electron baths in $\rho_B$ so that these quadratic coupling terms do not appear in
$H_{\text{int}}$ which only contains nonlinear electron-photon interactions. 

The standard diagrammatic expansion (or equation-of-motion method) can be used to cast the result in a Dyson equation,
\begin{align}
\begin{split}
D(z,\tau;z',\tau') &=  D_0(z,\tau;z',\tau') + \\
&\!\!\!\!\!\!\!\!\!\!\!\!\!\!\!\!\!\!\!\!\!\!\!\!\!\!\!\!\sum_j \int\!d\tau_1 \int\! d\tau_2 D_0(z,\tau;z_j,\tau_1) \Pi_j(\tau_1, \tau_2) D(z_j,\tau_2;z',\tau').
\end{split}
\end{align}
Because of the  extreme locality in our interaction terms, the self energies $\Pi_j$ take discrete values at the
site of electrons, and is diagonal in index $j$.   Applying the Langreth rules\cite{haug96}, and using time-translational invariance, the contour-ordered photon Green's function can be made simple in the frequency domain after Fourier transform.  This results in a pair of equations, the Dyson equation for the retarded component,
\begin{align}
\begin{split}
D^r(z,z',\omega) &= D_0^r(z,z',\omega)\, + \\
&\quad \sum_{j} D^r_0(z,z_j,\omega) \Pi^r_{j}(\omega) D^r(z_j,z',\omega),\label{photonDr}
\end{split}
\end{align} 
where $\Pi_{j}^r(\omega)$ is the retarded photon self energy of dot $j$, and the Keldysh equation which will be discussed in the next subsection.

In order to fully specify the problem and discuss its solution, one needs to give a recipe to compute $D_0^r$, which is
the Green's function for the free photon (including the effect of ``Rubin'' baths on the left and right sides) governed by the Hamiltonian $H_\gamma$.  
\begin{equation}
D_0^r(z,t;z',t')  = \frac{1}{i\hbar} \theta(t\!-\!t') \bigl\langle [ \phi(z,t), \phi(z',t') ]\bigr\rangle_{H_\gamma}.
\end{equation}
The easiest approach is to consider the equation of motion of the retarded Green's function.  Taking derivatives
with respect to $t$ twice, using the commutation relations for $\phi$ given by Eq.(\ref{eq-comm-phi}), we obtain
\begin{equation}
\epsilon_0 A \left(  \frac{1}{c^2} \frac{\partial^2\ }{\partial t^2} - \frac{\partial^2\ }{\partial z^2} 
\right) D_0^r(z,t; z',t') = \delta(z-z')\delta(t-t').
\end{equation}
Define the usual Fourier transform:
\begin{equation}
D_0^r( z,z',\omega) = \int_{-\infty}^{+\infty} D_0^r(z,t;z',0) 
e^{i\omega t}\, dt, 
\end{equation}
and using time translational invariance, we obtain
\begin{equation}
-\epsilon_0 A \left[ \left( \frac{\omega}{c} + i \eta\right)^2 + \frac{\partial^2\ }{\partial z^2} \right] D_0^r( z,z',\omega) = \delta(z-z').
\label{eq-D0r}
\end{equation}
We have added a damping term $\eta\to 0^+$ so that the inverse Fourier transform
satisfies $D_0^r(z,t,z',0) = 0$ for $t<0$, consistent with the definition. 
The differential equation can be solved, yielding the solution
\begin{equation}
D_0^r(z,z',\omega) = - \frac{e^{i(\frac{\omega}{c} + i \eta)|z-z'|}}{2\Omega}, \label{solutionGF}
\end{equation}
where $\Omega = i \epsilon_0 A \left( \frac{\omega}{c} + i \eta\right)$, and it is an important parameter which appears prevalently in this work.

\subsection{Solution of Dyson equation} \label{Solution_Dyson}

Although the full photon Green's function $D^r$ is defined for the continuum of $z$, its solution is essentially
characterized by the set of points ${z_j}$ where electrons sit.   Thus, we can solve the Dyson equation in matrix form,
specializing $z$ and $z'$ to the points $\{z_j\} \cup Z$, where $Z$ is a set of discrete points outside the locations
of the electrons.  This feature is very convenient, as it turns a continuum problem defined for all $z$ into a discrete
problem.  The solution is given in matrix form by $D = (D_0^{-1} - \Pi)^{-1}$. 

Alternatively, we can also act the differential operator appear{ing} on the left of $D_0^r$ in Eq.~(\ref{eq-D0r}) to the Dyson equation, and obtain the differential equation:
\begin{align}
\begin{split}
-\epsilon_0 A & \left[ \left( \frac{\omega}{c} + i \eta\right)^2 + \frac{\partial^2\ }{\partial z^2} \right]
D^r(z,z',\omega) = \\
& \delta(z-z') + 
 \sum_{j} \delta(z-z_j) \Pi^r_{j}(\omega) D^r(z_j,z',\omega).
\end{split}
\end{align} 
This equation can be interpreted as the scalar potential generated by a unit active (external) charge located {at}
$z'$, together with the induced extra charges at the electron sites, $z_j$, due to the linear response to the 
applied field.  Indeed, the induced charge at site $j$ is given by $\delta q_j = \Pi^r_j \phi(z_j)$, and $\Pi^r_j$ is the associated response function (dynamic susceptibility of the charge). 

The differential equation can be solved using transfer matrix\cite{Sipe:87}, or a straightforward boundary-condition matching.  Consider the case $z' < 0$, for example.  In each region
we can write {the solution} as forward-moving and backward-moving waves:
\begin{equation}
D^r(z,z') = \begin{cases}
b e^{-i\tilde{k}z}, & z < z',\\
c e^{i\tilde{k}z} + d e^{-i\tilde{k}z}, & z' < z< 0,\\
A e^{i\tilde{k}z} + B e^{-i\tilde{k}z}, & 0 < z< d,\\
t e^{i\tilde{k}z}, & d < z,
\end{cases}
\end{equation}
where we define the ``wavevector'' $\tilde{k} = \frac{\omega}{c} + i \eta$.  Since we demand
$D^r$ to be a bounded function at $|z| \to \infty$, for $z<z'$ ($z>d$) the wave moves only
backward (forward).  The function is continuous at the points $z'$, 0, and $d$ where we
have charges:
\begin{align}
\begin{split}
b \gamma' &= \frac{c}{\gamma'} + d \gamma', \\
c +d &= A + B, \\
A\lambda + \frac{B}{\lambda} &= t \lambda.
\end{split}
\end{align} 
We have defined the parameters $\lambda = e^{i\tilde{k}d}$,
$\gamma' = e^{i\tilde{k}|z'|}$.  The first derivatives need to be discontinuous so as
to generate the Dirac delta functions on the right-hand side of the equation.  This gives
\begin{align}
\begin{split}
\left(d \gamma' - \frac{c}{\gamma'}   - b \gamma'\right) \Omega  &= 1, \\
\bigl( B - A + c -d \bigr) \Omega &= (A+B)\Pi_0, \\
\left( A\lambda - \frac{B}{\lambda} - t \lambda \right) \Omega &= t \lambda \Pi_1.
\end{split}
\end{align} 
The six unknowns can be solved through these six linear equations.

For convenience of later calculations, we give here the solution of the Dyson equation for the two-dot case.
They also degenerate to a one-dot or no dot case if we set $\Pi_1 \equiv \Pi_1^r(\omega)$ or $\Pi_0$ or both to zero.  We note that the retarded Green's function is symmetric in space arguments, $D^r(z,z',\omega) = D^r(z',z,\omega)$.  The matrix elements associated with the locations of two dots are:
\begin{align} \label{Photon_int}
\begin{split}
D_{00} &=  D^r(0,0,\omega) = \frac{(1-\lambda^2) \Pi_1 + 2\Omega}{\mathcal{D}}, \\
D_{01} &=  D_{10} = D^r(0,d,\omega) = \frac{2\lambda \Omega}{\mathcal{D}}, \\
D_{11} &= { D^r(d,d,\omega) }  = \frac{(1-\lambda^2) \Pi_0 + 2 \Omega}{\mathcal{D}}, 
\end{split}
\end{align} 
where we have defined  
\begin{equation}
\mathcal{D} = (\lambda^2 - 1) \Pi_0 \Pi_1 - 2 \Omega\,(\Pi_0 + \Pi_1) - 4 \Omega^2.
\end{equation}
For $-L/2<z<0$, the rest of the elements can be expressed in terms of $D_{jk}$ ($j,k=0,1$).  We have 
\begin{align} \label{photon_retarded}
\begin{split}
D^r(z,-L/2,\omega) &=  \frac{(\gamma^2-1)}{2\gamma\Omega}\delta + \gamma \delta D_{00}, \\
D^r(z, 0,\omega) &=  \gamma D_{00}, \\
D^r(z,d,\omega) &=  \gamma D_{01}, \\
D^r(z,L/2,\omega) &=  \frac{\gamma\delta}{\lambda} D_{01},
\end{split}
\end{align} 
where we have defined $\gamma = e^{i \left( \frac{\omega}{c} + i \eta\right) |z|}$ and 
$\delta= e^{i \left( \frac{\omega}{c} + i \eta\right) L/2}$.  For $ 0 < z < d$, we obtain
\begin{align} \label{Photon_0_d}
\begin{split}
D^r( z, 0,\omega) &=  \left( \frac{\gamma^2-\lambda^2}{\gamma} \Pi_1 + 2 \gamma \Omega\right)\frac{1}{\mathcal{D}}, \\
D^r( z,d,\omega) &= \left((1-\gamma^2)\Pi_0 + 2 \Omega\right)\frac{\lambda}{ \gamma \mathcal{D}}{,} \\
D^r( z,-L/2,\omega) &=  {D^r( z, 0,\omega)}\,\delta , \\
D^r( z,L/2,\omega) &= {D^r( z,d,\omega)}\,\frac{\delta}{\lambda}.  
\end{split}
\end{align} 
For $d<z< L/2$, we have
\begin{align}
\begin{split}
D^r( z,-L/2,\omega) &=  \frac{\gamma\delta}{\lambda} D_{01}, \\
D^r( z, 0,\omega) &= \frac{\gamma}{\lambda} D_{01}, \\
D^r(z,d,\omega) &=  \frac{\gamma}{\lambda} D_{11}, \\
D^r( z,L/2,\omega) &= \frac{(\gamma^2-\lambda^2)\delta}{2\gamma\lambda^2\Omega} +  \frac{\gamma\delta}{\lambda^2} D_{11}.
\end{split}
\end{align}

\subsection{Keldysh equation}
While the retarded Green's functions describe the nature of wave propagation and the equation of motion, (nonequilibrium) thermal dynamic distributions are given by the lesser or greater Green's functions.   Since
the left side for $z<-L/2$ and $z>L/2$ are designated as photon baths, their effect should be reflected in the distribution.  Thus, the lesser Green's function must be based on the decoupled subsystems, symbolically, in the form
$D = d + d \Pi D$ (defined on contour), where $d$ is the Green's function of the isolated center (see Appendix~\ref{photon_bath}), $\Pi$ includes the contributions from the dots, $\Pi_i$, as well as the photon baths, $\Pi_L$ and $\Pi_R$.
The Keldysh equation associated with this contour-ordered Dyson equation is then
\begin{equation}
D^<(z,z',\omega) = \sum_j D^r(z,z_j,\omega) \Pi_j^<(\omega) D^a(z_j,z',\omega),\label{photonD<}
\end{equation}
where $\{ z_j \} = \{ -\frac{L}{2},0,d,\frac{L}{2} \}$ and $\{ j \} = \{ L,0,1,R \}$ for the two-dot model.
The advanced Green's function is given by $D^a(z,z',\omega) = D^r(z',z,\omega)^{*}$.

In Appendix~\ref{app:selfenergy}, we show that the bath self energy is:
\begin{eqnarray}
\Pi^<_\alpha(\omega) &=& 2\Omega N_\alpha(\omega) , \quad \alpha=L,R\\
\Pi^>_\alpha(\omega) &=& 2\Omega \bigl(1+N_\alpha(\omega)\bigr),
\end{eqnarray}
where $N_\alpha(\omega) = 1/\bigl(\exp(\beta_\alpha \hbar \omega) -1\bigr)$ is the Bose function at
temperature $T_\alpha = 1/(k_B\beta_\alpha)$.  Since our baths have negative-definite Hamiltonians, they have to be assigned
negative temperatures, $\beta_\alpha < 0$, for a convergent partition function (see Appendix~\ref{anti_commute}).

\section{Energy Currents}\label{sec:Currents}

After setting up the machinery to compute the Green's functions, we now consider how to connect them to physical
observables.  In our problem, the most important physical quantities are the energy currents.  We define
the energy leaving the baths as positive, thus, 
\begin{equation}
I_\alpha = - \left\langle \frac{dH_\alpha}{dt} \right\rangle, 
\end{equation}
where $\alpha =L$, $R$, and $j$.   We denote the left (right) photon baths by $H_L$ ($H_R$), while
$H_j^B$, $j=0,1,\cdots$ refer to the electron baths.  Referring to Fig.~\eqref{Two_dots}, the conservation
of energy is easy to understand, $I_L + I_0 + I_1 + I_R = 0$.   The lead currents are given by the well-known
Meir-Wingreen formulas\cite{haug96,MeirWingreen}:
\begin{subequations}  
\begin{align}  
I_{j}&= \int_{-\infty}^{+\infty} \frac{dE}{2\pi\hbar} E\, \bigl[ G_j^>(E) \Sigma_j^<(E) - G_j^<(E) \Sigma_j^>(E) \bigr], \\
I_{\alpha}&=- \int_{-\infty}^{+\infty} \frac{d\omega}{4\pi} \hbar \omega \,\bigl[D^>(\omega,z_\alpha,z_\alpha)\Pi^<_\alpha(\omega) - \nonumber \\
& \qquad\qquad\qquad\qquad\quad D^<(\omega,z_\alpha,z_\alpha)\Pi^>_\alpha(\omega) \bigr] ,
\label{eq:1C}
\end{align}
\end{subequations}
where $j=0,1$, $\alpha = L,R$, and $z_L = -L/2$, $z_R=L/2$. 

In addition to the currents leaving the baths, we can also ask: what is the energy current
between the dots, or between the dot and photon baths?  Towards that end, we shall derive an expression for the energy current carried by the scalar photons. For lack of a better name, we shall call it as the ``Poynting scalar'', in analogy to classical electrodynamics and reminding oneself that this energy is carried by scalar photons.  According to the Hamiltonian (treated classically),
Eq.~(\ref{eq:Hamiltonian}), the energy per unit volume is 
\begin{equation}
u(z,t) = -\frac{\epsilon_0}{2} \left[ \frac{\dot{\phi}^2}{c^2}+\left(\frac{\partial \phi}{\partial z}\right)^2 \right].
\end{equation}
Differentiating this expression with respect to time, and using the equation of motion of $\phi$ in vacuum, we
write $\partial u/\partial t + \partial j /\partial z = 0$, with
\begin{equation}
j(z,t) = \epsilon_0 \dot{\phi} \frac{ \partial \phi}{\partial z}.
\end{equation}
This will be our starting point to derive a quantized ``Poynting scalar'' at location $z$. First and foremost, a symmetrization is in place to ensure hermiticity: 
\begin{equation}
\epsilon_0\dot{\phi}\frac{ \partial \phi}{\partial z} \longrightarrow \frac{\epsilon_0}{2} \left[ \dot{\hat{\phi}} \frac{ \partial \hat{\phi}}{\partial z}+\frac{ \partial \hat{\phi}}{\partial z} \dot{\hat{\phi}} \right] \label{current_temp}
\end{equation}
We now discuss the necessity of anti-normal ordering to remove the zero-point motion contribution. Consider the Hamiltonian Eq.~\eqref{eq:Hamiltonian} \emph{without} the electron-photon interaction $H_{\textrm{int}}$, which is a solvable system, with the scalar field $\phi$ given by:
\begin{align}
\hat{\phi}(z,t) = \sum_q \sqrt{\frac{\hbar}{2\omega_q s L}} \left( \hat{a}_q e^{i(qz-\omega_q t)} + \rm{H.c.} \right).\label{eq:scalar}
\end{align}
Here, $q$ is the wavevector and $\omega_q = c|q|$ is the dispersion relation. The bosonic annihilation $\hat{a}_q$ and creation operator $\hat{a}_q^\dagger$ satisfy an unusual commutation relation, i.e. $[\hat{a}_q,\hat{a}_p^\dagger]=-\delta_{qp}$.

Were \eqref{current_temp} the quantized expression of the ``Poynting scalar'', one would have:
\begin{align}
\begin{split}
\hat{j}(z,t) =& \epsilon_0 \sum_{q,q'} \mathrm{sgn}(q') \frac{\hbar \sqrt{\omega_q\omega_{q'}}}{2 s c L} (\hat{a}_qe^{i(qz-\omega_q t)} - \rm{H.c.} \it) \\&\times (\hat{a}_{q'}e^{i(q'z-\omega_{q'} t)} - \rm{H.c.}) \\
=&\epsilon_0 \sum_{q,q'} \mathrm{sgn}(q') \frac{\hbar \sqrt{\omega_q\omega_{q'}} }{2 s c L} \Big( [ \hat{a}_q\hat{a}_{q'}e^{i(q+q')z-i(\omega_q+\omega_{q'})t}\\&- \hat{a}_q\hat{a}^\dag_{q'}e^{i(q-q')z-i(\omega_q-\omega_{q'})t}]+\textrm{H.c.} \Big),
\end{split}
\label{no_mo}
\end{align}
which, upon taking the expectation value with respect to the zero-photon state $|0 \rangle$, yields:
\begin{equation}
\langle 0|\hat{j}(z,t)|0\rangle = -\epsilon_0 \sum_q \frac{\hbar\omega_q}{2scL}\textrm{sgn}(q)\langle 0|\hat{a}_q^\dag \hat{a}_q+\hat{a}_q\hat{a}_q^\dag|0\rangle.
\end{equation}
Recall that in present there is no $H_{\textrm{int}}$ but only the free photon field, so the above is expected to vanish. In Appendix~\ref{anti_commute}, we show that the unusual commutation relation imposes $\hat{a}^\dag_q|0\rangle =0$. Therefore, one should perform an anti-normal ordering on \eqref{current_temp}:
\begin{equation} 
\hat{j}=\frac{\epsilon_0}{2} |: \dot{\hat{\phi}} \frac{ \partial \hat{\phi}}{\partial z}+\frac{ \partial \hat{\phi}}{\partial z} \dot{\hat{\phi}} :|, \label{current_g}
\end{equation} 
with $|:\dots :|$ the anti-normal ordering, so that in the non-interacting scenario the current expectation value is zero. Having achieved the purpose of quantizing the ``Poynting scalar'', we shall also drop the hats on the quantum operators. We show in Appendix C that when the expectation value is taken (with respect to a nonequilibrium steady state), the above can be written as:
\begin{align}    \label{Current_expression}
\langle j(z) \rangle  =& \epsilon_0  \int_0^{\infty} \frac{d\omega}{\pi} \hbar \omega  \mathrm{Re} \left. \frac{\partial D^>(\omega,z,z')}{\partial z'} \right|_{z'=z}  .
\end{align}

Energy conservation dictates that $I_L+I_0 =Aj(d/2)$ and $I_1+Aj(d/2) + I_R=0$. Therefore, it is mandatory for the ``Poynting scalar'' to agree with the heat current entering the photon bath. Indeed, it can be shown that $I_{L,R}= Aj(z)$ when $z<0$ or $z>d$, as outlined in the next section. In what follows, we shall refer to the integrand in Eq.~\eqref{Current_expression} as the ``spectral transfer function''.

\section{Currents in various limits}\label{sec:Currents_var_limit}
Notice that the parameters $\eta,L$ are introduced to dodge the pole of the retarded unperturbed photon Green's function, and to give our problem a finite size. Recall also that the speed of light $c$ should be taken as $\infty$ for a gauge-consistent theory. In this section, we consider different limiting procedures for these parameters $\eta,c,L$, and discuss their significance by studying the resulting current expression.   

\subsection{$\eta \to 0^+$ limit}
First, we consider the perfect medium limit, $\eta \to 0^+$, keeping the other parameters of
the model fixed.  Only in this limit, the energies $I_\alpha$ from various baths are strictly
conserved during wave propagation. Indeed, for $\eta\to 0^+$, the parameters $\lambda$, $\delta$, and $\gamma$ become phase factors with unit
modulus.  As a result, the retarded and lesser photon Green's functions become independent of the central region size $L$, and 
$j(z)$ is piecewise constant.  

We give the heat current calculated from Meir-Wingreen formula, $I_L$, $I_R$, as well as the  ``Poynting scalar'' expression  $Aj(z)$, for $z<d$, $0<z<d$, and $z>d$ without committing ourselves to the actual form of the self energies
due to electrons, $\Pi_{0,1}^{r,<,>}$.  For the bath spectral functions, we have
$\Pi_\alpha^r - \Pi_\alpha^a = \Pi_\alpha^> - \Pi_\alpha^< = 2\Omega$, $\alpha = L,R$. So these formulas are still fairly general.  The formulas are
obtained straightforwardly using the solutions of the retarded Green's function $D^r$, together
with the Keldysh equation, $D^{<,>} = D^r \Pi^{<,>} D^a$, and the assumption
$|\lambda| = |\gamma| = |\delta| = 1$:
\begin{align} 
\begin{split}
I_{L} &= -\int_{-\infty}^{+\infty}\!\! \frac{d\omega}{2\pi}\,  \hbar \omega\,\Omega \Big\{
|D_{00}|^2 \left( \Pi_0^> N_L - \Pi_0^< (N_L+1) \right) + \\
& |D_{01}|^2 \left( \Pi_1^> N_L - \Pi_1^< (N_L+1) + 2\Omega(N_R - N_L) \right) \Big\}.
\end{split}
\end{align} 
Here, $N_L$ and $N_R$ are the Bose functions associated with the photon baths, and
$D_{00}$ and $D_{01}$ are defined earlier in Sec.~\ref{Solution_Dyson}.
A similar expression can be written down for $I_R$.   The ``Poynting scalar'' formula is
\begin{align} 
\begin{split}
Aj(z<0) &= -\int_{0}^{+\infty}\!\! \frac{d\omega}{\pi}\,  \hbar \omega\,\Omega \Big\{
|D_{00}|^2 \left( \Pi_0^> +  2\Omega (N_L+1) \right)  \\&+
|D_{01}|^2 \left( \Pi_1^> + 2\Omega (N_L+1) \right) + \\
&\quad (N_L+1)\bigl( D_{00}^{*}- D_{00} \bigr) \Big\}.
\end{split}
\end{align} 
Since $I_L$ and $Aj(z<0)$ have the same physical meaning--the energy from the left bath
going to the right--they should be identical.  Indeed, this can be shown using an important identity for the Green's functions\cite{datta}:
\begin{equation}
D^r - D^a  = D^r( \Pi^r - \Pi^a) D^a,
\end{equation}
where $D^r$, $\Pi^r$, etc., are $4\times 4$ matrices whose entries correspond to the positions $z\in \{-L/2,0,d,L/2\}$, and $\Pi^r$
is diagonal with diagonal elements $\{\Pi_L^r, \Pi_0, \Pi_1, \Pi_R^r\}$.  Using this identity, 
and the relation $\Pi_j - \Pi_j^{*} = \Pi_j^> - \Pi_j^<$, $j=0,1$,  
the last term ${\rm Im}\,D_{00}$ in the ``Poynting scalar'' formula can be transformed into a form
involving the self energies, and the equivalence is proved.

\subsubsection*{Local equilibrium approximation (LEA)}
As the electron contributions to photon self energies, $\Pi_j^{r,<}$, are not known exactly, one has
to make various approximations in order to make concrete predictions.  In this paper, we consider three kinds of approximations.  The local equilibrium approximation, the Born approximation (BA), and the self-consistent Born approximation.  We first elaborate on the LEA.  By LEA, we mean that the electrons are in 
respective equilibrium with the associated electron baths.  Specifically, we assume that the 
electron-photon interaction is so weak that the thermal equilibria of the electrons
are not disturbed.   Thus, the charge's degrees of freedom being in equilibrium, the corresponding self energies 
satisfy the fluctuation-dissipation theorem:
\begin{align} 
\begin{split}
\Pi_j^< &= N_j (\Pi_j - \Pi_j^{*}) , \quad j=0,1\\
\Pi_j^> &= (N_j+1) (\Pi_j - \Pi_j^{*}) . \\
\end{split}
\end{align} 
Notice that the photon self energies are related to the charge's degrees of freedom $c_j^\dagger c_j$, so the flucutation-dissipation theorem is boson-like, with Bose function $N_j = 
1/(\exp(\beta_j \hbar \omega) - 1)$.

Under LEA, our theory is identical to Rytov fluctuational electrodynamics\cite{yap16} (except here, we 
consider charge fluctuations).
We denote the transmission function by $T_{\alpha \beta}(\omega)$. For our two-dot model, there is a ``detailed-balance'' condition, i.e. $T_{\alpha \beta}(\omega) = T_{\beta \alpha}(\omega) $. Based on such relation, the
four-terminal Laudauer-$\rm  B\ddot{u}ttiker$ form of current expression can be derived for the bath $\alpha = L$, 0, 1, and $R$: 
\begin{align} \label{LB_formula}
\begin{split}
I_{\alpha} = \int_0^\infty \frac{d\omega}{2\pi}\,  \hbar \omega\!\!\! \sum_{\gamma={L,0,1,R}}\!\!\! \bigl[ N_\alpha(\omega) - N_\gamma(\omega) \bigr] T_{\alpha \gamma}(\omega).
\end{split}
\end{align} 
This form guarantees energy conservation explicitly, $\sum_\alpha {I_\alpha} = 0$. The transmission functions are given explicitly by: 
\begin{align} 
T_{01} &= \left|\frac{4\Omega}{\mathcal{D}}\right|^2 {\rm Im} \Pi_0\,{\rm Im} \Pi_1,\label{transmission01} \\
T_{0L} &=  4|\Omega| \bigl| D_{00}\bigr|^2\,{\rm Im} \Pi_0,\label{transmission0L}\\
T_{0R} &= \frac{16}{|\mathcal{D}|^2} |\Omega|^3\,{\rm Im} \Pi_0, \\
T_{1L} &= \frac{16}{|\mathcal{D}|^2} |\Omega|^3\,{\rm Im} \Pi_1,\\
T_{1R} &= 4|\Omega| \left| D_{11} \right|^2 {\rm Im} \Pi_1,\label{transmission1R} \\
T_{LR} &= \frac{(2\Omega)^4}{|\mathcal{D}|^2}.\label{transmissionLR}
\end{align} 
The diagonal terms $T_{\alpha\alpha}$ are irrelevant, and other terms are obtained by symmetry,
$T_{\alpha\beta} = T_{\beta\alpha}$. 

We consider some special cases.  If $\Pi_0 = \Pi_1 = 0$, i.e. a system with no dot, then all the
transmission functions are 0 except $T_{LR} = 1$.  This represents a perfect transmission from
the left bath to right bath. If the two photon baths have different temperatures, the whole system is not in thermal equilibrium and an energy current flows between the two photon baths.
\begin{figure}[htp] 
  \centering
  \includegraphics[totalheight=60mm]{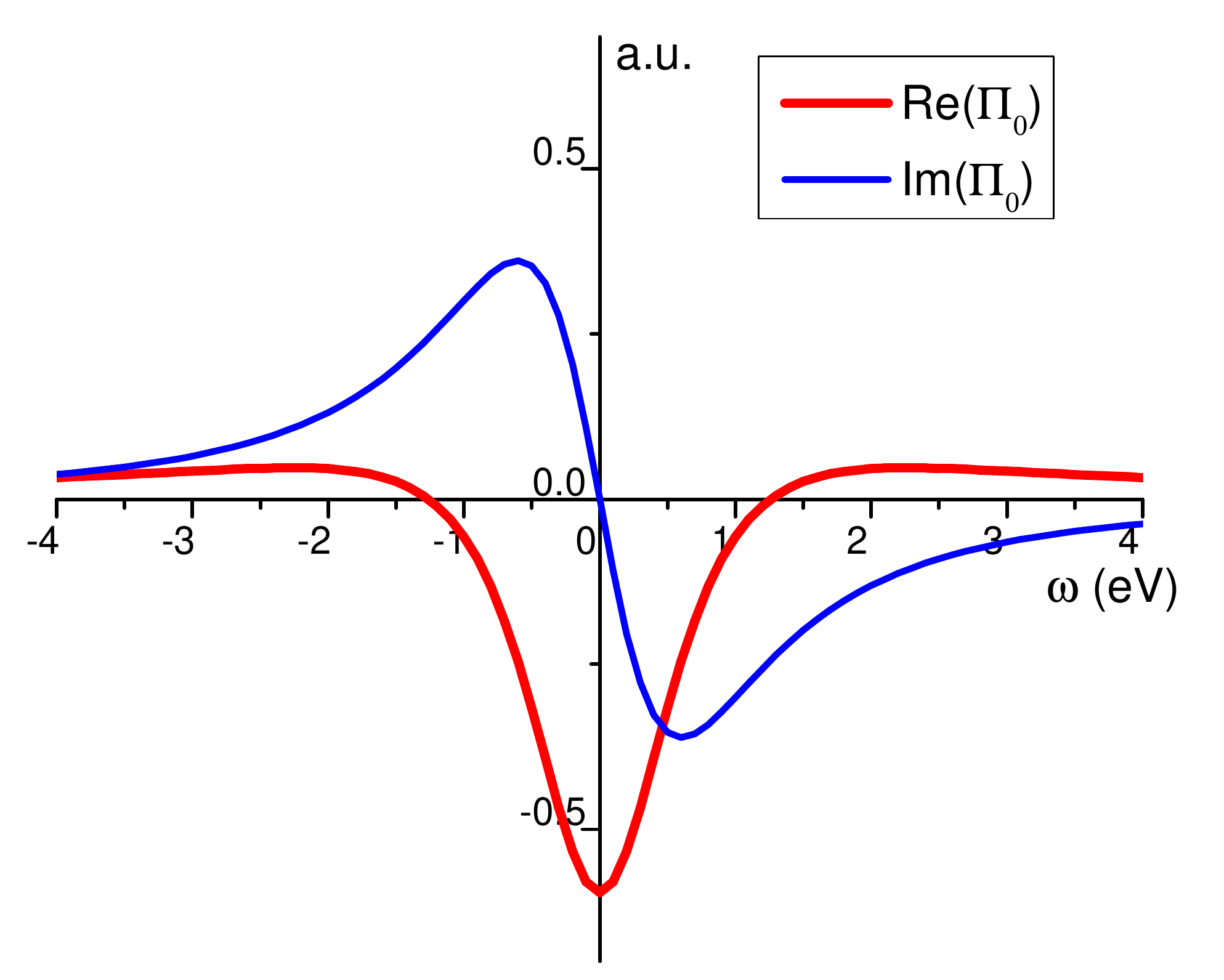}
  \caption{Real and imaginary part of the photon self energy.}
\label{ImPi}
\end{figure}

Next, we consider the energy exchange between scalar photons and electrons. Notice that ${\rm Im} \Pi_j < 0$ if $\omega >0 $ as shown in Fig.~\ref{ImPi}. This means that the transmission coefficients between the photon baths and electron, i.e. Eqs.~\eqref{transmission0L}--\eqref{transmission1R} are negative. Now, scalar photon baths, having negative-definite Hamiltonians, must be assigned a negative temperature (see Appendix~\ref{anti_commute}). Therefore, the difference of the Bose function between a scalar photon and an electron bath is always negative, i.e. $N_{L/R}-N_{0/1}<0$ for $\omega>0$. Thus the integrand in the Landauer-B\"{u}ttiker formula \eqref{LB_formula} is always positive when we consider the energy contributed by scalar photons to the electrons. This matches the expectation that a negative temperature is hotter than any positive temperature\cite{PhysRev.103.20,swendsen2016negative}. However, as we discuss in the following subsection, scalar photon baths will not play any role when $c\to\infty$.

\subsection{$c \to \infty$ limit}
Infinite speed of light is the limit that must be taken for a gauge-consistent theory. One could think of $c$ as a speed that is needed only for field quantization. Once it has served this purpose, in all resulting expressions, e.g. the current formula, transmission coefficients, one eliminates $c$ by letting it goes to infinity.

Returning to the transmission coefficients, observe that the speed of light appears  in $\Omega = i \epsilon_0 A \omega/c = O(1/c)$ (we already sent $\eta$ to $0^+$), and
also $\mathcal{D} = O(1/c)$. Using this fact, the coefficients that involve at least one photon bath, i.e. Eqs.~\eqref{transmission0L}--\eqref{transmissionLR} can be shown to vanish, leaving $T_{01}$ as the only nonzero transmission.  This shows that in the strict Coulomb interaction limit,
energy cannot be transmitted to long distances.   We also comment that the infinite speed
limit is a robust limit independent of how the other limits ($\eta$ and $L$) are taken. 

When $c\rightarrow \infty$, the transmission function between dot 0 and dot 1, Eq.~\eqref{transmission01} simplifies to:
\begin{equation} \label{Transmission_c_inf}
T_{01} = \frac{4\, {\rm Im}\Pi_0\, {\rm Im} \Pi_1}{ |\Pi_0 + \Pi_1 - \Pi_0 \Pi_1/C|^2}, 
\end{equation}
where the parameter $C = \epsilon_0 A/d$ is precisely the capacitance of the parallel plate capacitor.  This shows that our model contains indeed the physics of parallel-plate capacitors.  Furthermore, the pole of \eqref{Transmission_c_inf} gives a critical length $\tilde{d}$:
\begin{equation}
\tilde d = \epsilon_0 A \left( \frac{1}{\Pi_0(0)} + \frac{1}{\Pi_1(0)} \right),
\end{equation} 
where the self energies are evaluated at zero frequency.  Notice that $\tilde d$ is a negative quantity
in normal systems, so the transmission never really diverges.  But $\tilde d$ is a length scale that differs 
markedly from the thermal wavelength $\lambda_T = 2\pi\hbar c/(k_B T)$.

\subsection{$L \to \infty$ limit}
In contrast to the above two limits, we get rid of the photon bath in a technically consistent way, that is, we let the medium (the vacuum) be dissipative, and
we put the baths at $\pm \infty$. 
In the limit $L \to \infty$ but keeping $\eta$ finite, the medium consumes the bath energies.  As a result, the bath
contribution disappears.  Since the baths are infinitely far away, their energies are dissipated on the way to
the dots, so it is equivalent to having a distribution which is strictly 0.  We are left with the dots only.

\begin{figure}[H]  
  \centering
  \includegraphics[totalheight=60mm]{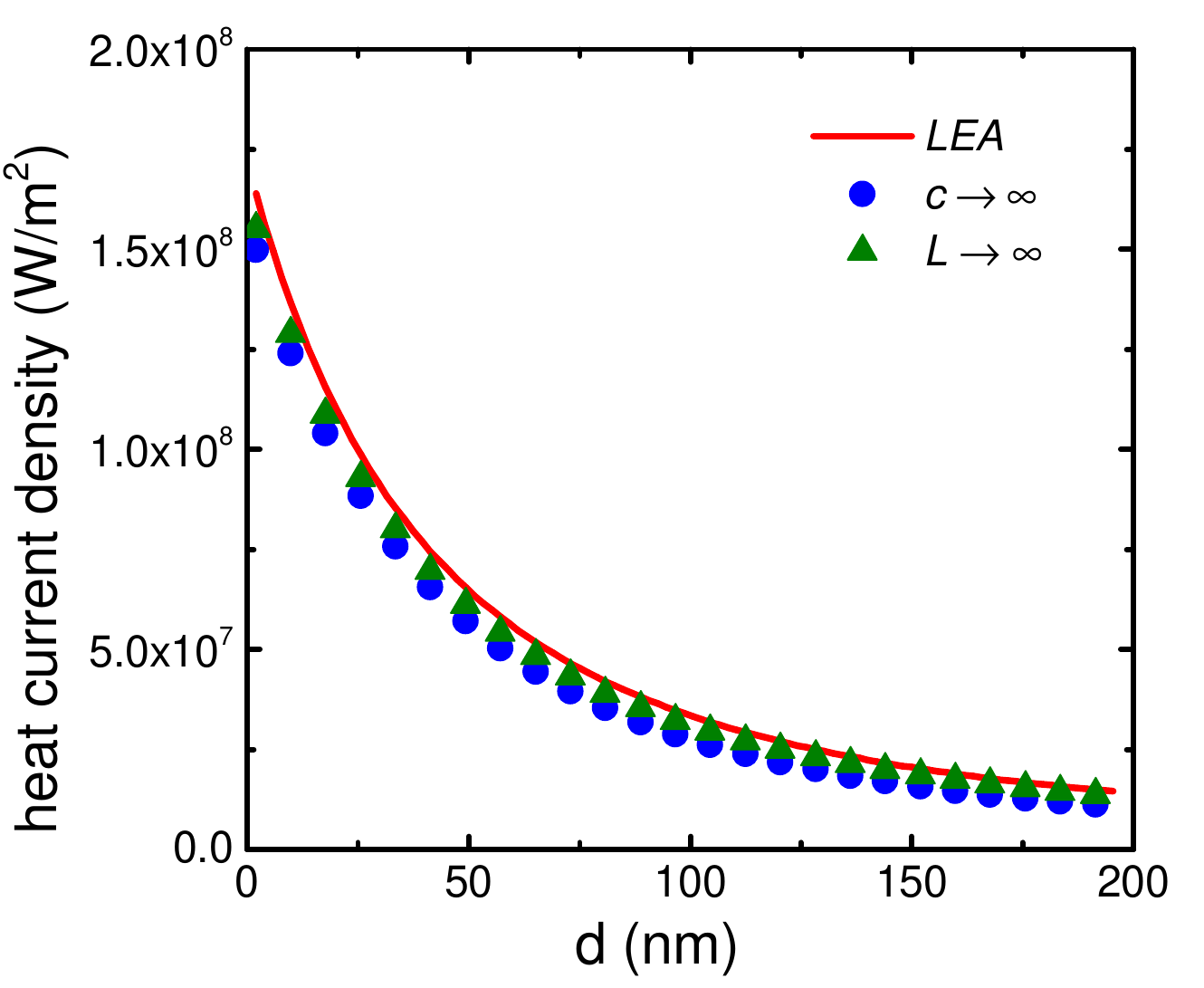}
  \caption{Distance dependence of heat current under different limits. The temperature of dot 0 is 1000 K and dot 1 is 300 K. The onsite energy of dot 0 is 0.0 eV and dot 1 is 0.01 eV. The chemical potential of dot 0 is 0.0 eV and dot 1 is 0.02 eV. We take the wide band limit, with $\Gamma_0=1$ eV and $\Gamma_1 = 0.5$ eV, $E_0=E_1=50$ eV. The $Q$ factor is 1$e$ and the area of capacitor $A=$ 19.2$\times$19.2 $\mathrm{nm}^2$.}
   \label{diff_limit}
\end{figure}

In Fig.~\ref{diff_limit}, we numerically check the distance dependence of heat current in three different limits. We find that blue circles and green triangles almost coincide with the red solid line. This indicates that the currents in three different limits are consistent: the $c\rightarrow \infty$ limit is similar with both LEA and BA. However, energy conservation of the whole system is not guaranteed under BA. Hence, in numerical calculations, it is the self-consistent Born approximation (SCBA) that we use to calculate the nonequilibrium ``Poynting scalar''. 

\section{Calculations of self energies}\label{sec:Self_energy}

To make numerical predictions of our analytic formulas for the currents, we need
a way of computing the photon self energies.  If the local equilibrium approximation (LEA) 
is assumed, we only need to know the retarded self energies.  This correlation
function can be calculated exactly since we assume that the electrons are kept in equilibra with the baths, unaffected by the photons.  A formula is given in Appendix \eqref{Matsubara}. 
However, if 
the electron-photon interaction of our model is taken into account, there is no exact method for such calculations, except the purely formal Hedin equation\cite{hedin65}.  Thus, we consider various approximation schemes.    
\subsection{Photon self energy}
\begin{figure}[H]
	\centering
	\includegraphics[width=0.95\columnwidth]{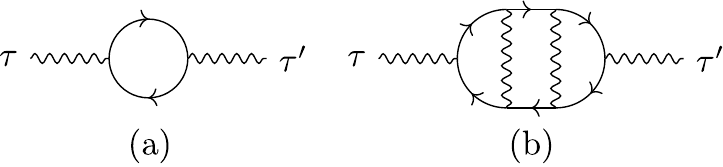}
	\caption{Diagrammatic expansion of the photon Green's function \eqref{photonD}, including the photon lines at both ends. (a) Lowest-order non-trivial loop diagram. (b) Typical ladder diagram in higher-order expansions which are not included.}
	\label{Ferman}
\end{figure}

In a diagrammatic expansion of the contour-ordered scalar photon Green's function
$D(\tau, \tau')$, we can write the result as a Dyson equation.  The lowest order 
term in this expansion for the self energy (all the irreducible diagrams) $\Pi(\tau, \tau')$ 
is \cite{mahan00,stefanucci}
\begin{equation}
\Pi_{jk}(\tau, \tau') \approx \frac{1}{i\hbar} \bigl\langle 
T_\tau q_j(\tau) q_k(\tau') \bigr\rangle_{H_e}, \label{loopdiagram}
\end{equation}
where $q_j = (-Q)c_j^\dagger c_j$ is the charge operator at site $j$. Notice that $\Pi_{jk}(\tau, \tau') = 0$ if $j \neq k$, true for all orders, because we excluded inter-dot coupling in the Hamiltonian \eqref{eq:Hamiltonian}. In other words, electrons are not allowed to jump from one site to the other.  Since $\Pi$ is diagonal, we use
the notation $\Pi_j \equiv \Pi_{jj}$. 
After applying Wick's theorem\cite{fetter2003quantum}, the resulting expression can be written as a series of diagrams, see Fig.~\ref{Ferman}. 
If we stop here and use the equilibrium
distributions of the electrons $G_0$, we obtain LEA, as discussed earlier. This is however not
an exact result.  We have neglected many diagrams and kept only the loop (polarization) diagram (see Fig.~\ref{Ferman}(a)).  In particular, we have ignored
a class of nontrivial diagrams known as ladder diagrams (see Fig.~\ref{Ferman}(b)).  In order to include systematically
the ladder diagrams, we need to solve the Bethe-Salpeter equation\cite{fetter2003quantum,altland}, which 
is computationally much more involved.  
The next level of approximation is the Born approximation (BA).  In the Born approximation, we
do not use the equilibrium Green's functions of the electrons.  Instead, we include the lowest order
nonlinear self energy as a perturbation.  As it turns out, this is a bad approximation, because energy
conservation is not done consistently. More precisely, the energy conservation, $\sum I_\alpha = 0$,
is violated at the next order ($Q^4$) of approximation. 

Thus, for numerical simulations, we adopt the well-established self-consistent Born approximation (SCBA)\cite{lu2007}. This means that, in the lowest
order self energy expansion, we replace the unperturbed Green's
function $G_0$ and $D_0$ (i.e. the Green's functions when $H_{\mathrm{int}}=0$) by the 
interacting ones $G$ and $D$ for both the photon self energy $\Pi$ as well as the electron
self energy $\Sigma$.  This calculation requires iterations until a convergence criterion is attained. 

As mentioned, we only keep diagram (a) in Fig.~\ref{Ferman}. The nonlinear photon self energy in contour time is given by:
\begin{align}
\begin{split}
 \Pi_{j}(\tau,\tau') = -i\hbar Q^2 G_{j}(\tau,\tau') G_{j}(\tau',\tau).
\end{split}
\end{align} 
After applying the Langreth rules\cite{haug96}, the photon self energies in real time read:
\begin{align}
\begin{split}
\Pi_{j}^>(t) =& -i\hbar Q^2 G_{j}^>(t) G_{j}^<(-t),\\
\Pi_{j}^<(t) =& -i\hbar Q^2 G_{j}^<(t) G_{j}^>(-t),\\
\Pi_{j}^r(t) =& \theta(t) \bigl[\Pi_{j}^>(t) - \Pi_{j}^<(t)\bigr],
\end{split}\label{photonPi}
\end{align}
where $\theta(t)$ is the Heaviside step function. In numerical calculations, we transform Eq.~\eqref{photonPi} to the frequency domain using the fast Fourier transform.


\subsection{Electron self energy}

\begin{figure}[H]
	\centering
	\includegraphics[scale=1.1]{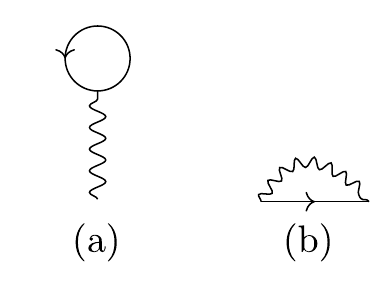}
	\caption{Self energies in the diagrammatic expansion of the electron Green's function $G(\tau, \tau')$. (a) Hartree term. (b) Fock term. }
	\label{electron}
\end{figure}

For an unperturbed electron, the retarded Green's function is written in standard form,
\begin{equation}
G^r_{0,j}(\omega) = \frac{1}{\hbar \omega -v_j - \Sigma^r_{b,j}(\omega)},
\end{equation} 
where the subscript $0$ refers to the absence of interaction and $v_j$ is the onsite energy of dot $j$, $\Sigma^r_{b,j}(\omega)$ is the retarded electron bath self energy for electron bath $j$, given explicitly by:
\begin{equation}
\Sigma^r_{b,j}(\omega) = \sum_{k\in\textrm{bath}} \frac{|V_{jk}|^2}{\hbar\omega+i\eta_k-\epsilon_{jk}},
\end{equation}
where one must then specify the mode energy $\epsilon_{jk}$, coupling strength $V_{jk}$ and damping factor $\eta_k$ for mode $k$ in bath $j$. In this work, we use the Lorentz-Drude model\cite{dattaNEGF}:
\begin{equation}
\Sigma^r_{b,j}(\omega)=\frac{\Gamma_j/2}{i+ \hbar \omega/E_{0,j}}.
\end{equation}
The wide-band limit is obtained by taking $E_0 \to \infty$, which reduces the self energy to a constant.  Adding a decay makes it more physical, as the real time function
$\Sigma^r(t)$ has an exponential decay with decay time $\hbar/E_0$. This model has
been used for electronic transport studies of quantum dots\cite{dattaNEGF,haug96}. 
In calculations, we set $\Gamma$ and $E_0>0$ as constant. The free electron lesser Green's function can be easily derived using the fluctuation-dissipation theorem for electrons. When electron-photon interaction is active, the full retarded Green's functions are obtained from a Dyson equation and the lesser Green's function can be derived from Keldysh equation with external interaction self energy:
\begin{align}
G^{r}_j(\omega) =& \frac{1}{\hbar \omega - v_j - \Sigma_{b,j}^r(\omega)- \Sigma^{r}_{n,j}(\omega)},\label{electronGr}\\
G^{<}_j(\omega)=& G^r_j(\omega) [ \Sigma^{<}_{b,j}(\omega) + \Sigma^<_{n,j}(\omega) ] G^a_j(\omega).\label{electronG<}
\end{align} 
$\Sigma_{n,j}^{r,<}(\omega)$ is the retarded and lesser self energy due to nonlinear electron-photon interactions at site $j$. $\Sigma_{b,j}^<(\omega)$ is the lesser Green's function of the electron bath which can be derived from the fluctuation-dissipation relation, i.e. $\Sigma^<_{b,j}(\omega) = -f_j(\omega) [\Sigma^r_{b,j}(\omega) -\Sigma^a_{b,j}(\omega) ] $, where $f_j(\omega)= 
1/\bigl\{\exp([\beta_j(\hbar \omega - \mu_j)]+1\bigr\}$ is the Fermi distribution for dot $j$.  $G^r_j(\omega)$ is the total retarded Green's function and $G^<_j(\omega)$ is the total lesser Green's function.  For our two-dot capacitor model, the nonlinear self energies $\Sigma^{r,<}_{n,j}(\omega)$ can be obtained from the Hartree and Fock diagrams\cite{zhanglifa2013} (see Fig.~\ref{electron}) in a diagrammatic 
expansion for the Green's function $G(\tau, \tau')$.
For each of the dots $j=0$, 1, up to first non-trivial diagrams, we have, for the nonlinear electron self energies:
\begin{align}
\begin{split}
\Sigma_{n,j}^<&(t) = i\hbar Q^2 G^<_{j}(t)D^<(z_j,z_j,t),\\
\Sigma_{n,j}^r&(t) = \theta(t)[\Sigma^>_{n,j}(t)-\Sigma^<_{n,j}(t)]\\&-i\hbar Q^2\sum_{k=0,1}\Big[-G^<_k(t=0)\int dt' D^r(z_j,z_k,t')\Big] .\label{electronSigma}
\end{split}
\end{align}
Hence, electrons interact with photons via the photon Green's function $D$, which in turn interacts with the electron Green's function $G$. Thus the equations have to be solved iteratively.

\subsection{Computational Details}
In actual calculations, we always take the limit $c \to \infty$, under which Eq.~\eqref{Photon_int} reduces to:
\begin{align} \label{photon_GF_cinf}
\begin{split}
D_{00} =& \frac{C - \Pi_{1}}{\mathcal{D}}, \\
D_{11} =& \frac{C - \Pi_{0}}{\mathcal{D}}, \\
D_{01} =& D_{10} = \frac{C}{\mathcal{D}}, \\
\end{split}
\end{align} 
where, as before, $C={A\epsilon_0}/{d} $ is the capacitance but the denominator is now $\mathcal{D} = \Pi_{0} \Pi_{1} - C(\Pi_{0}+\Pi_{1})$. With these coefficients, the radiative heat current in the limit $c\to \infty$ simplifies from Eq.~\eqref{Current_expression} to:
\begin{align} \label{Current_c_inf}
\begin{split}
\langle j \rangle = \frac{1}{A} \int_0^\infty \frac{d\omega}{\pi} \mathrm{Re} \left[{i\hbar \omega}{|D_{10}|^2}(\Pi_1^> \mathrm{Im}\Pi_0 - \Pi_0^>  \mathrm{Im}\Pi_1)  \right].
\end{split}
\end{align} 
Numerically, the terms \eqref{photon_GF_cinf} diverge in the high frequency regime, leading to unstable iterations under SCBA. To overcome this difficulty, we multiply Eq.~\eqref{photon_GF_cinf}  with an artificial damping factor $\frac{1}{1+({E}/{E_{d}})^2}$ to help with convergence under SCBA and $E_d$ is set as 4 eV in the following  calculations. We have checked numerically that despite this artificial damping, energy conservation is still satisfied after several iterations.


\section{Results and discussion} \label{RESULT}

\begin{figure}[htp]  
  \centering
  \includegraphics[totalheight=60mm]{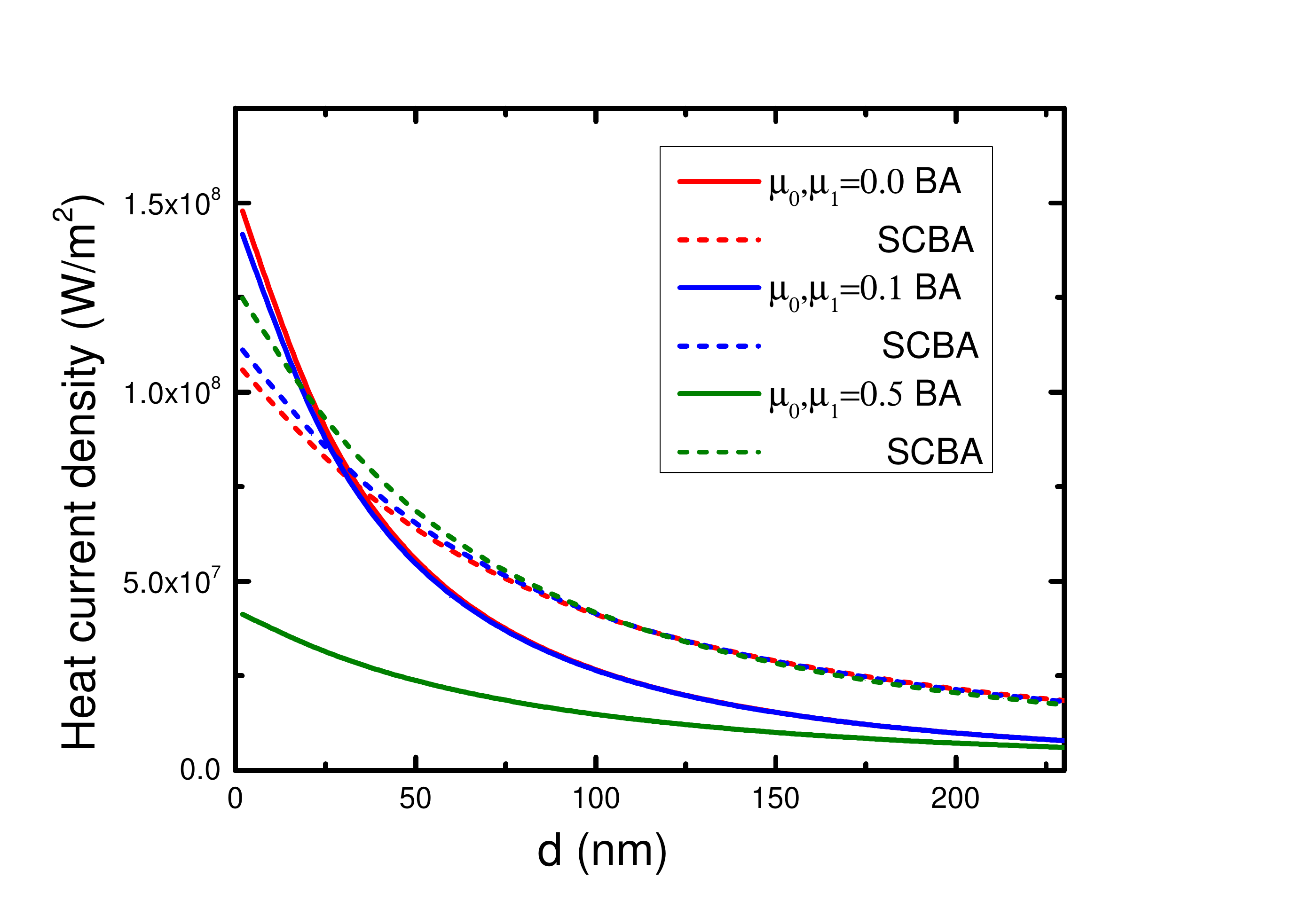}
  \caption{Distance dependence of heat current density for different dot chemical potentials, using either BA or SCBA. The temperature of dot 0 is 1000 K and dot 1 is 300 K. The chemical potentials of both dots are equal. $Q=1e$.}
   \label{BAvsSCBA}
\end{figure}

\paragraph{BA vs SCBA}
In this paragraph, we first discuss the differences between BA and SCBA. The solid lines in Fig.~\ref{BAvsSCBA} show the heat current between two quantum dots under BA with different dot chemical potentials. A first observation is that the chemical potential of quantum dots hugely influences the heat current. As the chemical potential increases, the total heat current under BA decreases and eventually converges to the same value at large dot separation. Such convergence can be understood by the evanescent properties of the scalar field: with an increasing gap between the dots, the heat current density decreases and ultimately becomes insensitive to the source properties. However, in sharp contrast to BA, currents computed under SCBA (dotted lines) increase with chemical potential. This can be understood as follows. As the chemical potential is increased (within a reasonable parameter region), the dot becomes highly occupied, leading to a stronger electron-photon interaction. Under SCBA, the self energies of electrons and scalar photons are updated at each iteration, allowing this strong interaction to be more aptly captured. On the other hand, BA crudely stops at just the first iteration. Therefore, one expects the heat current to be enhanced under SCBA and not necessarily so for the case of BA. Another observation is that both BA and SCBA show a convergence (at large dot separation) of heat currents towards two different chemical potentials. Finally, it is seen that the chemical potential only affects the heat current at small distances, which again can be understood from the rapidly-decaying property of the scalar field.  


\paragraph{$Q$ and area dependence}
\begin{figure}[htp]  
  \centering
  \includegraphics[totalheight=110mm]{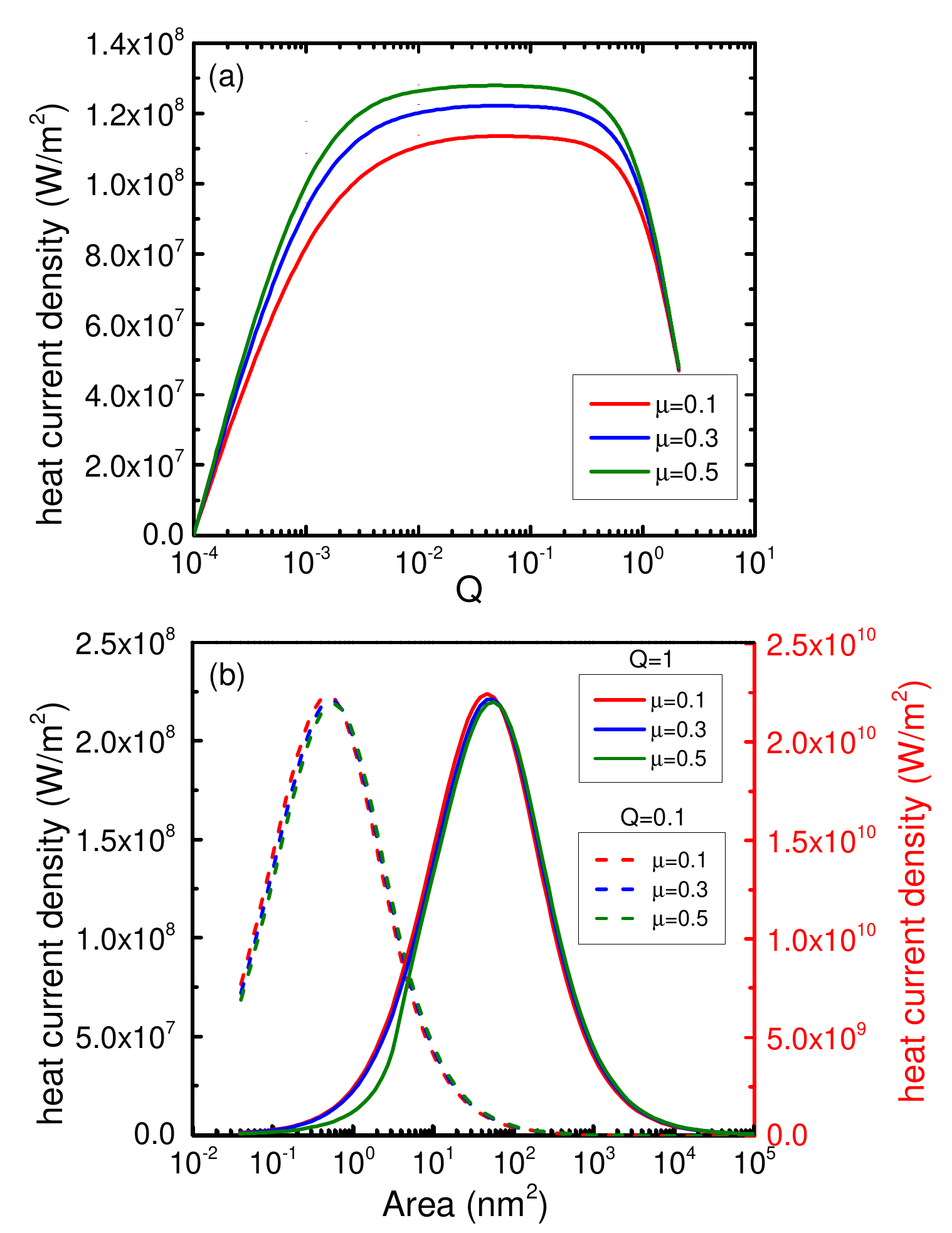}
  \caption{Charge-dependence and area-dependence of the heat current density for different dot chemical potentials under SCBA. (a) Charge-dependence of current density. (b) Area dependence of current density. The left axis scale corresponds to the solid line and right axis scale corresponds to the dashed line. The temperature of dot 0 is 1000 K and dot 1 is 300 K. The chemical potentials of both dots are the same. The distance between the two dots is 19.2 nm.}
   \label{Q_dependent}
\end{figure}

 In our model, the parameter $Q$ is the maximum charge at the dot. It also determines the strength of electron-photon interaction. Fig.~(\ref{Q_dependent}a) shows the $Q$ dependence of the heat current density at different chemical potentials. As in Fig.~\eqref{BAvsSCBA}, the heat current between two quantum dots can be amplified by increasing the dot chemical potential. Besides, the heat currents at different chemical potentials converge to the same value as $Q$ is increased. This can be understood as follows: with a strong electron-photon interaction, the electrons in quantum dots are less prone to excitations by thermal fluctuation alone. Hence, the effect of chemical potential on the heat current is expected to diminish for large $Q$. 

For smaller values of $Q$, the heat current is more easily controlled by the chemical potential of quantum dots. However, as $Q$ approaches zero, the heat current decays rapidly, and vanishes for $Q=0$. Therefore, this parameter needs to be optimally chosen to control the heat transfer. {For gold, the free electron concentration is $5.90\times 10^{22}$~cm$^{-3}$ and the estimated value of $Q$ is of the order of $10^3e$. In this case, the scalar field is therefore not expected to be the dominant heat transfer channel. On the contrary, semiconductors or certain 2D materials such as graphene have much lower free electron density compared to bulk gold, $Q\approx 10^{-1}e$, hinting at a more controllable RHT by tuning the chemical potential.} 

Fig.~(\ref{Q_dependent}b) shows the area dependence of the heat current density between the two dots. Comparing the peak positions of the solid and the dashed line (corresponding to different $Q$), we find that the peak position is proportional to $Q^2$. Hence, within this parameter regime ($0.1e<Q<1e$), the dot area $A$, which governs partly the capacitance $C=\epsilon_0A/d$, plays a crucial role in the heat transfer.


\paragraph{Scalar field vs blackbody limit}

\begin{figure}[htp]  
  \centering
  \includegraphics[totalheight=60mm]{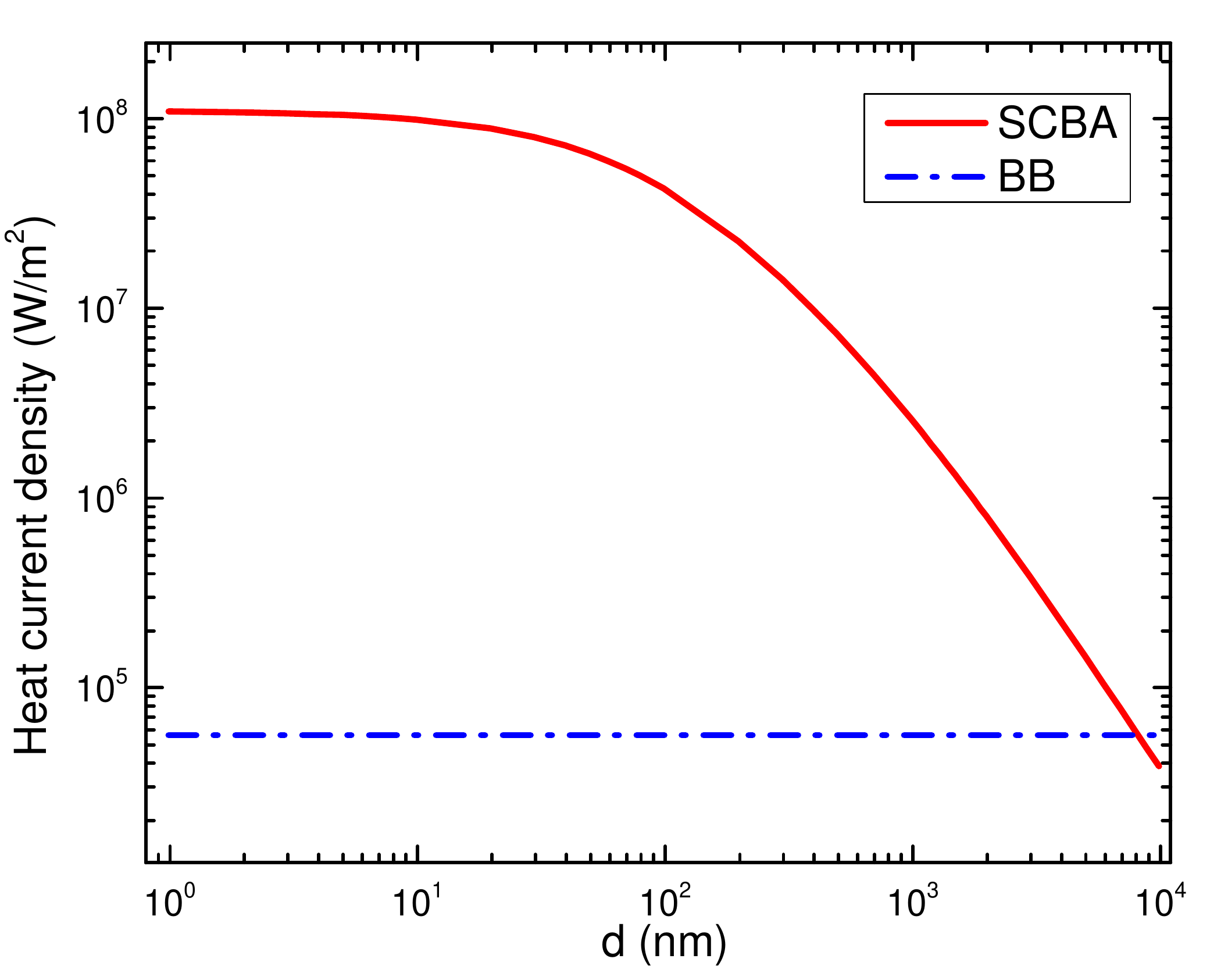}
  \caption{Current calculation under SCBA and the blackbody limit in log-log plot. The temperature of dot 0 is 1000 K and dot 1 is 300 K. The chemical potential of dot 0 is 0.0 eV and dot 1 is 0.02 eV. $Q=1e$ }
   \label{vsBB}
\end{figure}

Fig.~\ref{vsBB} presents  the distance dependence of heat current density in a double logarithmic plot. {For large distances, the heat current density decreases like $d^{-2}$. Such a scaling law arises from the capacitor property, which can be manifested in the expression for transmission coefficient when $c \rightarrow \infty$, i.e. Eq.~\eqref{Transmission_c_inf}. This is different from the case of p-polarized waves, which diverges like $d^{-2}$ only for short distance ($< 0.1$ nm). Our prediction of a $d^{-2}$ scaling at large distances for a nanocapacitor} can be experimentally tested, such as with a tip-plane heat current measurement. However, the heat current is suppressed at close separation because of the large value of capacitance when the two dots are nearly in contact. With such large capacitance, our calculation predicts a constant transmission in the $d\rightarrow 0$ limit. 

Finally, Fig.~\eqref{vsBB} shows an extremely large enhancement of heat transfer mediated by the scalar photons. Comparing the red solid and blue dashed-dotted lines at 10 nm, we find that the heat current density is two thousand times larger than the blackbody limit. This result demonstrates that the heat transfer channel provided by electron-photon interaction is the dominant one for nanocapacitors at small separations.

\paragraph{Scalar-field-based thermal rectification}
Thermal rectification is a phenomenon in which the heat flux depends on the sign of the temperature difference of two bodies. In the literature, a vast majority of the previous works focuses on geometry or phonon-induced rectification\cite{terraneo2002controlling,chang2006solid,li2012colloquium,zhang2010ballistic,yang2009thermal}. While not much has been done on voltage-controlled thermal rectification, our simple two-dot model is suitably tailored for that. Therefore, we discuss in this paragraph the tunability of thermal rectification via chemical potential. To quantify the strength of thermal rectification, we define the rectification constant:
\begin{align}
\begin{split}
R= \frac{J_{0\rightarrow 1} - J_{1\rightarrow 0}}{J_{1\rightarrow 0}},
\end{split}
\end{align} 
where the temperature difference has the same magnitude but opposite sign to $J_{0\rightarrow 1}$ and $J_{1\rightarrow 0}$.
\begin{figure}[htp]  
  \centering
  \includegraphics[totalheight=120mm]{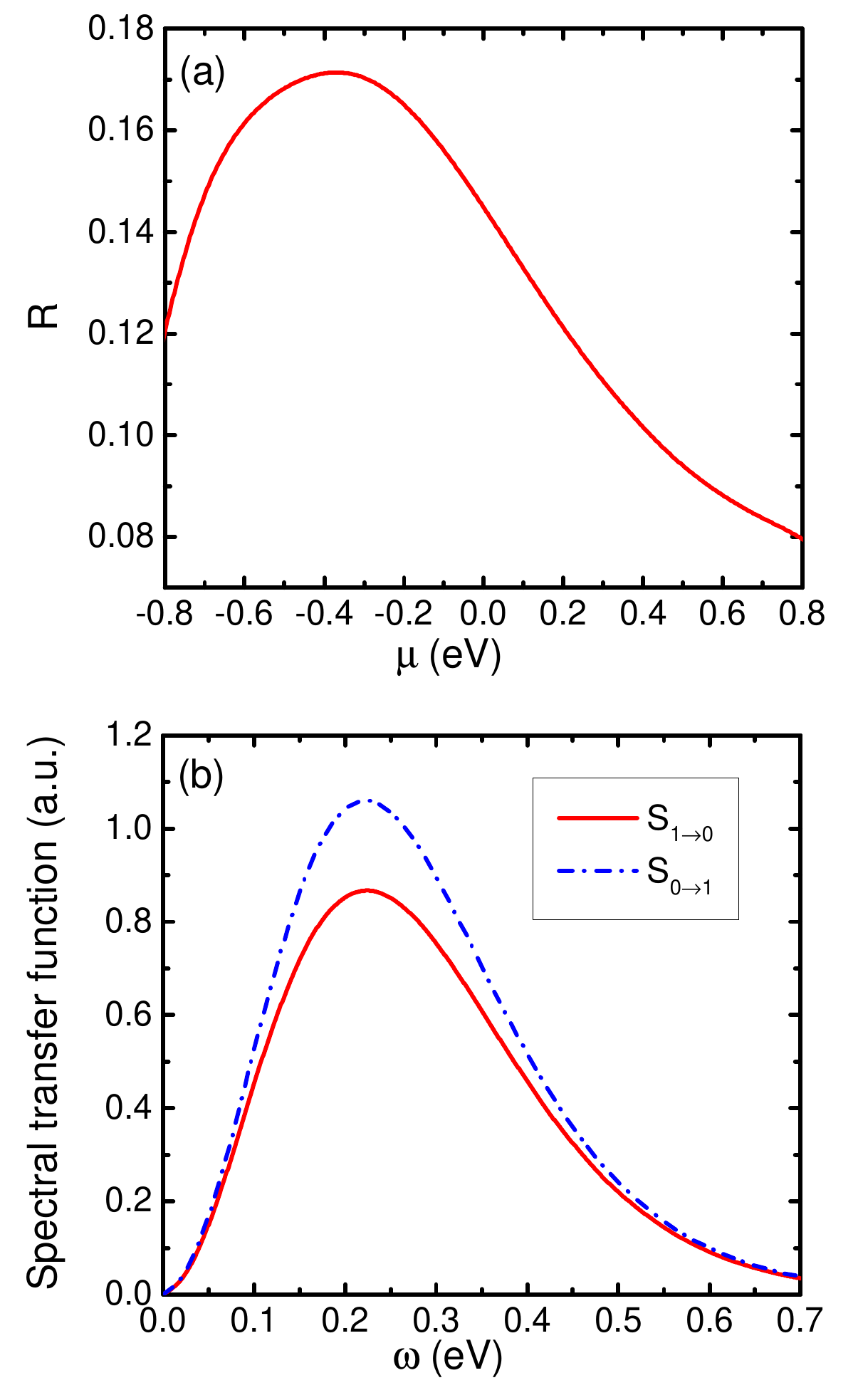}
  \caption{Thermal rectification effects. (a) Rectification constant as a function of the chemical potential of the quantum dots, with both dots having the same chemical potential. (b) Spectral transfer function of two quantum dots with switched temperature. The chemical potential of two quantum are set as $-$0.4 eV. In above two figures, the temperatures are set at 1000 K and 300 K. Distance between the two quantum dots is set as 19.2 nm. $Q=1e$. }
   \label{REC}
\end{figure}

Fig.~\ref{REC} shows the chemical potential dependence of rectification strength. For simplicity, we set the same chemical potential for both dots. Within experimentally-accessible parameter regimes, i.e. $-$0.8 eV to 0.8 eV, the rectification constant varies from 0.08 to 0.17, and is maximized at $-$0.4 eV. On the other hand, the heat flux is very small when the chemical potential is greater than 1.0 eV or smaller than $-$1.0 eV. Such parameter regime corresponds to a highly occupied or unoccupied electron level, both indicating the lack of carriers for heat flux generation. One can also study rectification by means of the spectral transfer function. In Fig.~\ref{REC}(b), the blue dash-dot ($S_{0\rightarrow 1}$) and red solid line ($S_{1\rightarrow 0}$) have the same peak position at 0.22 eV but different maxima. In the wide-band limit for the electron bath, the value $\Gamma$, which determines the dot-bath coupling, is set at 1 eV for dot 0 and 0.5 eV for dot 1. Both parameters are in the strong-coupling regime, so the energy flux from electron bath to the dot depends greatly on the coupling strength. This asymmetric coupling is what drives the rectification effect.

\section{Conclusion}\label{sec:Conclusion}
In this work, using the method of NEGF, we studied the heat flux between a nanoscale capacitor, mediated by the Coulomb interaction, generated by charge fluctuation. By modeling the capacitor as a double quantum dot, we ruled out transverse electrical current and focused only on charge fluctuation. Keeping in mind that $c\rightarrow\infty$ must be taken at the end for a gauge-invariant theory, we worked under Lorenz gauge to quantize the scalar potential.  We then outlined the framework of NEGF and solved the Dyson equation. From a classical continuity equation, we derived a Coulomb heat flux, dubbed ``Poynting scalar'', discussed its quantization, and related it to the greater Green's function. Next, we studied the current expression under various limits of three model parameters. This was achieved by focusing on the local equilibrium approximation, which bridges our approach and Rytov's theory. To illustrate the greater generality of NEGF, we discussed two approximation schemes on the self energies: the BA and SCBA. Finally, we performed numerical simulations, explored the dependences of heat current on different parameters, and demonstrated how chemical potential can be used to tune radiative thermal rectification. We also found a thousandsfold enhancement of the heat current compared with the blackbody limit and discovered a new distance dependence ($1/d^2$) at large distances. 

Although our model might be unrealistic, it contains an essential feature of most thermal transport problems: a left/right partition of the whole system. It also helps address several issues, such as the ambiguity of heat fluxes (``Poynting scalar'' and photon bath energy current), the necessity of a negative definite scalar photon Hamiltonian, and perhaps most importantly how our approach is a generalization of Rytov's phenomenological theory. A more realistic problem, such as the heat transfer between two large graphene sheets placed parallel to each other, and a more interesting three-dot transistor model, will be the subjects of future work.

\section*{Acknowledgements}
This work is supported by FRC grant R-144-000-343-112.

\appendix

\section{Inverted harmonic oscillator} \label{anti_commute}
The photon baths can be understood as a collection of independent oscillators.  Thus,
it is sufficient to study a single-mode oscillator at, say, frequency $\omega_0>0$.  However, these oscillators are inverted,
with a negative kinetic energy and negative potential energy.  Such model has been studied
by Glauber\cite{glauber} as a quantum amplifying device.  Here we see it is necessary to
`build' a photon bath. 

Consider thus a single-oscillator Lagrangian $\mathcal{L}=T-V = -\frac{1}{2} (\dot{u})^2 - \left( -\frac{1}{2} \omega_0^2 u^2\right)$.  The conjugate momentum is $p= -\dot{u}$.  This implies
the canonical commutation relation is $\bigl[ u,p\bigr] = [\dot{u},u] = i \hbar$.  The
Hamiltonian is the negative of the usual one, $H= p\dot{u} - \mathcal{L} = 
-\frac{1}{2} \left(p^2 + \omega_0^2 u^2\right)$, which is negative definite. 
We can introduce the creation and annihilation operators in the usual way,
$u = \sqrt{\hbar/(2\omega_0)} \left( a + a^\dag\right)$, and 
$-p = \dot{u} = i\sqrt{\hbar\omega_0/2}  \left( -a + a^\dagger\right)$.  This leads to the
commutation relation:
\begin{equation}
\bigl[a, a^\dagger \bigr] = -1.
\end{equation}
The Hamiltonian can then be written as:
\begin{equation}
H = - \frac{1}{2} \hbar \omega_0 \left( a a^\dagger + a^\dagger a\right) = - \hbar \omega_0 \left(a a^\dagger + \frac{1}{2} \right),
\end{equation}
where the extra $-\hbar \omega_0/2$ is the zero-point motion energy, and $a a^\dagger$ is the number
operator.  Since $\bigl[ H, a\bigr] = - \hbar \omega_0 a$, and
$\bigl[ H, a^\dagger\bigr] =  \hbar \omega_0 a^\dagger$, the meaning of $a^\dagger$ ($a$)
increasing (decreasing) the energy by $\hbar \omega_0$ remains the same as in the usual oscillator.  Since
the eigen-energies of the system cannot be positive, the raising operation must terminate at the zero-number state. This gives the condition $a^\dagger | 0\rangle = 0$.  Thus, how the eigenstates are built
is now different.  The eigenvalues are $E_n = - \hbar \omega_0 (n+1/2)$, $n=0$, 1, 2, $\cdots$,
with the eigenvectors $\propto a^n|0\rangle$.  

We can work out the statistical mechanics of the problem.  The average occupation number
in a canonical ensemble is: 
\begin{align}
\begin{split}
\bigl\langle a a^\dagger \bigr\rangle_H &= \frac{ {\rm Tr}\left( a a^\dagger e^{-\beta H} \right)}{{\rm Tr}\left( e^{-\beta H} \right)} \\
&= \frac{1}{e^{-\beta \hbar \omega_0} - 1} \equiv N_{-\beta}(\omega_0).
\end{split}
\end{align} 
In deriving the above, one encounters a geometric series in $e^{\beta\hbar\omega_0}$, which, for $\omega_0>0$, converges only if $\beta<0$. Thus, we see that the partition function is defined only for $\beta<0$, i.e. statistical mechanics demands that scalar photon baths yield negative absolute temperature.  Also,
$\bigl\langle a^\dagger a \bigr\rangle_H = \bigl\langle a a^\dagger \bigr\rangle_H + 1
= - N_\beta(\omega_0)$.  One can check that the fluctuation-dissipation theorem holds in
the usual way,
\begin{align}
\begin{split}
D^<(\omega) &= N_\beta(\omega) \bigl( D^r(\omega) - D^a(\omega) \bigr), \\
D^>(\omega) &= \bigl(N_\beta(\omega)+1\bigr) \bigl( D^r(\omega) - D^a(\omega)\bigr), 
\end{split}
\end{align} 
where the Green's functions are defined in the variable $u$ according to the usual
convention\cite{wang08review,wang14rev}.

\section{Scalar photon bath} \label{photon_bath}
We discuss here the bath spectrum of the scalar photon. As per standard procedures of open quantum systems, we need to partition the photon field into system and baths. Our strategy is to first discretize the photon to have recourse to the well-developed surface Green's function\cite{jswpre07,wang08review}. We then take the continuum limit to obtain a bath spectrum of the photon field defined on a continuum.   
\subsection{Discretization of scalar photon field}
In Sec.~\ref{sec:THEORY}, when describing the photon Hamiltonian $H_\gamma$, we split the real line into
three regions for the left bath in $(-\infty, -L/2]$, the central system in $[-L/2,L/2]$ and the right bath in $[L/2,\infty)$.   However, given that the Hamiltonian
is an integral on the line, it is not clear what the interaction between the bath
and the central region is.  For this reason, we consider a discretized version of the model by
putting the problem on a 1D lattice with lattice constant $a$. The spatial derivative in potential energy thus becomes a finite difference, and the $z$ integral becomes a discrete sum:
\begin{equation}
\begin{split}
H_\gamma =& -s \int dz \frac{1}{2} \left[ \dot{\phi}^2+c^2\left(\frac{\partial \phi}{\partial z}\right)^2 \right] \\
\longrightarrow& -\frac{s}{2}\sum_n a\left[ \dot{\phi}_n^2 +c^2 \left(\frac{\phi_{n+1}-\phi_n}{a}\right)^2  \right].
\end{split} 
\end{equation}
Similar to phonon on a lattice, we set $x=j\cdot a$ and $\phi_j = \frac{1}{\sqrt{sa}} u_j$. Continuum limit is recovered for $a\rightarrow 0$. Putting a ``spring constant'' $k= \frac{c^2}{a^2}$, we find a discretized photon Hamiltonian\cite{jswpre07}:
\begin{align}
\begin{split}
H_0 = &-\sum_n \frac{1}{2} \dot{u}_n^2 - \frac{1}{2} k \sum_n (u_{n+1} - u_n)^2,
\end{split}
\end{align} 
which has the same form as a phonon Hamiltonian (except for the minus sign). For the equation of motion of $u_n$ we have:
\begin{align}
\begin{split}
\ddot{u}_n = k(u_{n+1} - 2u_n + u_{n-1}), \label{B5}
\end{split}
\end{align} 
which also follows from a direct discretization of the wave equation $\frac{\partial^2 \phi}{\partial t^2} = c^2 \frac{\partial \phi}{\partial x^2}$. Eq.~\eqref{B5} can be solved by setting: 
\begin{align}
\begin{split}
u_n =& A \lambda^n e^{-i\omega t}.
\end{split}
\end{align} 
Putting $\omega \rightarrow \omega + i\tilde{\eta}$ (for regularization) and $\lambda = e^{i q\cdot a}$, we find the dispersion relation for the discretized photon field: 
\begin{align}
\begin{split}
(\omega+i\tilde{\eta})^2  =& 2k [1- \cos(q\cdot a)].
\end{split}
\end{align} 
In the continuum limit $a\rightarrow 0$, the above reads $\omega+i\tilde{\eta} = \pm c\cdot q$, recovering the free photon dispersion relation in continuum. 

Using the discretized photon quadrature operator $u_j$, we define the decoupled retarded Green's function as:
\begin{align}
\begin{split}
\check{d}^r_{jk} (\omega) =  -\frac{i}{\hbar}  \int_0^\infty dt e^{i \omega t}\langle [ u_j(t), u_k(0) ] \rangle_{H_0} , \label{discreteGF}
\end{split}
\end{align} 
where $H_0$ is the unperturbed free photon Hamiltonian. Its relation to the usual one defined on a continuum, namely:
\begin{align}
\begin{split}
d^r (x,x',\omega)= -\frac{i}{\hbar} \int_0^{\infty} dt e^{i\omega t}  \langle [ \phi(x,t) , \phi(x',0) ] \rangle_{H_0}  , 
\end{split}
\end{align} 
 is given by:
\begin{align}
\begin{split}
d^r (x,x',\omega) = \lim_{a\rightarrow 0} \frac{1}{sa} \check{d}_{jk}^r (\omega) ,\label{eq:disc_cont}
\end{split}
\end{align} 
where $j\cdot a$ and $k \cdot a$ correspond respectively to $x$ and $x'$.

\subsection{Self energy of scalar photon bath}\label{app:selfenergy}
Consider a semi-infinite chain of discretized photons, labeled by $j=-1,-2,\cdots$. We derive the surface Green's function of this system by attaching a new site $j=0$ to the right-most site $j=-1$. The retarded Green's function \eqref{discreteGF}, when regarded as an infinite matrix, satisfies the following equations:
\begin{equation}
[(\omega + i\tilde{\eta})^2 - \widetilde{K}]\check{d}^r = -{I}, \label{surfaceGF}
\end{equation}where ${I}$ is the identity matrix (the minus sign originates from the unusual commutation relation $[u_j,\dot{u}_k] = -\delta_{jk}$), and $\widetilde{K}$ is an infinite tridiagonal matrix given by:
\begin{align}
\widetilde{K}=
\left[                 
  \begin{matrix}
    2k & -k     &       &   \\  
    -k & 2k     & -k&   \\  
      & -k & 2k & \ddots \\
      &       & \ddots     & \ddots \\
  \end{matrix}
\right]                .
\end{align} 
Focusing on the first column of the right-hand side of \eqref{surfaceGF}, we find a system of difference equations:
\begin{equation}
\begin{split} 
k \check{d}^r_{j-1,0}+[(\omega+i\tilde{\eta})^2-2k]\check{d}^r_{j,0}&+k\check{d}^r_{j+1,0} =0,\\ & j=-1,-2,\cdots \label{difference_eq}
\end{split}
\end{equation}
with boundary condition:
\begin{equation}
[(\omega+i\tilde{\eta})^2-2k]\check{d}^r_{0,0} + k\check{d}^r_{-1,0} = -1.\label{boundary_cond}
\end{equation}
To solve \eqref{difference_eq} and \eqref{boundary_cond}, consider the ansatz $\check{d}^r_{j,0}=\alpha\lambda^j$, where $j=0,-1,-2,\cdots$, which gives the quadratic equation:
\begin{equation}
k\lambda^{-1}+[(\omega+i\tilde{\eta})^2-2k]+k\lambda = 0. \label{eq:quadratic}
\end{equation}
This equation admits two roots: $\lambda_<,\lambda_>$ with $|\lambda_<|< 1$ and $|\lambda_>|>1$. We choose the second root so that $\check{d}^r_{j,0}$ goes to zero for $j\to -\infty$. Using the boundary condition \eqref{boundary_cond}, we find $\alpha=1/(k\lambda_>)$ and hence $\check{d}^r_{j,0}=\lambda_>^{j-1}/k$.   

\begin{figure}[H]
\centering 
\includegraphics[scale=1.25]{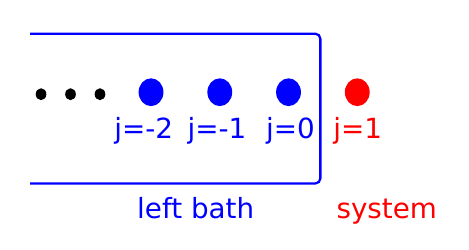}
\caption{Derivation of discretized photon bath self energy. Attaching the first system site $j=1$ to the semi-infinite left bath $j=0,-1,-2,\cdots$ leads to the identification of bath self energy, Eq.~\eqref{eq:selfenergybath}, from the Dyson equation describing the coupling.}\label{fig:surfGF}
\end{figure}

Next, we consider attaching the first system site $j=1$ to the semi-infinite chain (Fig.~\ref{fig:surfGF}).
This coupling can be described by a Dyson equation:
\begin{equation}
\begin{split}
\begin{bmatrix}
\check{D}^r_{11} & \check{D}^r_{01}\\
\check{D}^r_{10} & \check{D}^r_{00}
\end{bmatrix}=
\begin{bmatrix}
\check{d}^r_{11} & 0 \\
0 & \check{d}^r_{00}
\end{bmatrix}+\begin{bmatrix}
\check{d}^r_{11} & 0 \\
0 & \check{d}^r_{00}
\end{bmatrix}\begin{bmatrix}
0 & k \\ k & 0
\end{bmatrix}\begin{bmatrix}
\check{D}^r_{11} & \check{D}^r_{01}\\
\check{D}^r_{10} & \check{D}^r_{00}
\end{bmatrix},
\end{split}
\end{equation}
where the $\check{D}^r$'s denote the discretized retarded photon Green's functions that result from coupling the left bath to site $j=1$. From the equation for $\check{D}^r_{11}$, one identifies the left photon bath self energy $\check{\Pi}^r_L$:
\begin{equation}
\check{D}^r_{11} = [( \check{d}^r_{11} )^{-1}-\underbrace{(k\check{d}^r_{00}k)}_{\check{\Pi}^r_L}   ]^{-1}. \label{eq:selfenergybath}
\end{equation}
Hence, we find:
\begin{equation}
\begin{split}
\check{\Pi}^r_{L} &= k\check{d}^r_{00}k \\
&= k \lambda_<,
\end{split} 
\end{equation}
where we used the fact that $\lambda_> \lambda_<=1$ in view of the quadratic equation \eqref{eq:quadratic}, which can be solved (up to second order in $a$) by:
\begin{equation}
\begin{split}
\lambda_< &= e^{i(\frac{\omega}{c}+i\tilde{\eta})a}, \\
\lambda_> &= e^{-i(\frac{\omega}{c}+i\tilde{\eta})a}.
\end{split} 
\end{equation}
Armed with the retarded left photon bath self energy, we find the lesser self energy using the fluctuation-dissipation relation:
\begin{equation}
\begin{split}
\check{\Pi}^<_L &= N_L [\check{\Pi}^r_L-\check{\Pi}^a_L] \\
&\approx  N_L k\left[2ia\frac{\omega}{c}\right].
\end{split} 
\end{equation}
Finally, using the relation between discretized and continuum Green's functions, Eq.~\eqref{eq:disc_cont}, we obtain for the left photon bath:
\begin{equation}
\begin{split}
\Pi^<_L &= \lim_{a\to 0} sa \check{\Pi}^<_L \\
&= 2isc\omega N_L \\
&= 2 \Omega N_L,
\end{split} 
\end{equation}
where $\Omega=isc^2(\frac{\omega}{c}+i\eta)=i\epsilon_0 A(\frac{\omega}{c}+i\eta)$ as in the main text, and a damping factor $\eta$ has been added. 
\subsection{Meir-Wingreen formula for the scalar photon bath}
Lastly, we comment on the validity of the usual Meir-Wingreen formula. Since we have in hand a negative definite Hamiltonian for the scalar photon, a sign-flip is apparently needed compared to the usual case of positive definite bath Hamiltonian. However, a similar sign flip occurs for the Green's functions $D,\Pi$. Thanks to the symmetrical combination of the Green's functions ($D^<\Pi^>$ or $D^> \Pi^<$) in the integrand for current, the minus sign is canceled, allowing the usual Meir-Wingreen formula to remain intact without any sign-flip needed.



\section{Current expression}
Here we relate the quantized ``Poynting scalar'', Eq.~\eqref{current_g} to the greater Green's function as in Eq.~\eqref{Current_expression}. We first perform a nonequilibrium steady-state average on \eqref{current_g}, which amounts to:
\begin{equation}
\begin{split}
\braket{ j(z,0)} &= \frac{\epsilon_0}{2}\frac{\partial}{\partial t}\frac{\partial}{\partial z'}\Big[ \braket{|:\phi(z,t)\phi(z',0)+\textrm{H.c.}:|}    \Big]_{\substack{t=0\\ z'=z}}.
\end{split}
\end{equation}
In frequency domain, this yields:
\begin{equation}
\begin{split}
\braket{j(z)} &= \frac{-i\epsilon_0}{2}\int_{\mathbb{R}}\frac{d\omega}{2\pi}\omega \frac{\partial}{\partial z'} \braket{|:\phi(z,\omega)\phi(z',0):|}_{z'=z}\\&+\textrm{c.c.}  \label{integralcurrent}
\end{split} 
\end{equation}
where the second term inside the braket is still in time domain: $\phi(z',0)=\phi(z',t'=0)$.
We now split the scalar field into its positive frequency part $\phi^+(z,t)=\sum_q \sqrt{\frac{\hbar}{2\omega_qsL}}a_q(t) e^{iqz}$ and its negative frequency part $\phi^-(z,t) = [\phi^+(z,t)]^\dagger$, so that $\phi = \phi^+ +\phi^-$. In the following we outline the steps that lead to the conclusion $\braket{\phi^+\phi^+}=0=\braket{\phi^-\phi^-}$. 

Define a 2$\times$2 contour-ordered full Green's function $\bm{D}$, whose components are given by: $[\bm{D}]_{\sigma\sigma' }=-\frac{i}{\hbar}\braket{T_\tau\phi^\sigma\phi^{\sigma'}}$, where $\sigma,\sigma'\in\{+,-\}$. This splitting yields a Dyson equation: $\bm{D}=\bm{D_0}+\bm{D_0}\bm{\Pi}\bm{D}$, where the photon self energy $\bm{\Pi}$ is index-independent. For non-interacting Green's functions $\bm{D_0}$, the average is with respect to the photon thermal state. Therefore, an unbalanced number of creation and annihilation operator gives zero expectation value, i.e. one has $D_0^{++}=0=D_0^{--}$. Therefore, this identity extends to real-time components, e.g. retarded, advanced, lesser, greater. Next, we argue that in the frequency domain, the retarded Green's functions satisfy:
\begin{equation} 
\begin{alignedat}{2}
&(D^{++})^r = 0, 
&& \quad\quad(D^{--})^r = 0, \\
&(D^{+-})^r \propto \theta(\omega), 
&&\quad \quad(D^{-+})^r \propto \theta(-\omega).
\end{alignedat}
\end{equation}To this end, we first show that the non-interacting retarded Green's functions satisfy $(D^{+-}_0)^r\propto\theta(\omega)$ and $(D^{-+}_0)^r\propto \theta(-\omega)$. This is achieved by solving the non-interacting lesser Green's function $(D^{+-}_0)^<$. From the definition of scalar field \eqref{eq:scalar}, with the fact that $\langle a^\dag_{q}a_{q'} \rangle_0 = N(c|q|) \delta_{q,q'}$, where $N$ is the Bose function, passing the discrete sum to an integral, $\frac{2\pi}{L}\sum_q \rightarrow \int dq$, one finds, for $\omega\neq 0$:
\begin{equation}
(D_0^{+-})^<(z,\omega)=\frac{N(\omega)}{\Omega}\cos\left(\frac{\omega}{c}z\right)\theta(\omega).\label{C4}
\end{equation}
Now, non-interacting Green's functions obey the fluctuation-dissipation relation:
\begin{equation}
(D^{+-}_0)^<(z,\omega) = N(\omega)\{(D^{+-}_0)^r(z,\omega)-[(D^{+-}_0)^r]^*(z,\omega)  \}.
\end{equation}
From \eqref{C4}, the left-hand side is non-zero only if $\omega>0$, hence it must also be the case for the right-hand side, i.e.:
\begin{equation}
\textrm{Im}(D_0^{+-})^r(z,\omega) = -\frac{1}{2sc\omega}\cos\left(\frac{\omega}{c}z\right)\theta(\omega).
\end{equation}
Next, taking into account the unusual commutation relation, we find in a similar manner to \eqref{C4}:
\begin{equation}
(D^{-+}_0)^<(z,\omega)=-\frac{N(-\omega)-1}{\Omega}\cos\left(\frac{\omega}{c}z\right)\theta(-\omega).
\end{equation}
Applying again the fluctuation-dissipation relation for $D^{-+}_0$, we are led to conclude:
\begin{equation}
\textrm{Im}(D^{-+}_0)^r(z,\omega)=-\frac{1}{2sc\omega}\cos\left(\frac{\omega}{c}z\right)\theta(-\omega).
\end{equation}
Since we know exactly the unsplit non-interacting retarded Green's function, Eq.~\eqref{solutionGF}:
$D_0^r(z,\omega)=-e^{i\frac{\omega}{c}|z|}/(2\Omega)$, and that it splits according to: $D_0^r(z,\omega)=(D^{+-}_0)^r(z,\omega)+(D^{-+}_0)^r(z,\omega)$, we must have:
\begin{equation}
\begin{split} 
(D_0^{+-})^r(z,\omega)&= D^r_0(z,\omega)\theta(\omega),\\
(D_0^{-+})^r(z,\omega)&=D^r_0(z,\omega)\theta(-\omega).
\end{split}
\end{equation}
We now expand the matrix Dyson equation for $[\bm{D^r}]_{\sigma\sigma'}$:
\begin{align}
(D^{++})^r &= (D_0^{+-})^r\Pi^r\big[(D^{++})^r+(D^{-+})^r\big],\label{D++}\\
(D^{+-})^r &= (D^{+-}_0)^r\big[  1+\Pi^r(D^{+-})^r+\Pi^r(D^{--})^r \big],\label{D+-}\\
(D^{-+})^r &= (D^{-+}_0)^r\big[1+\Pi^r(D^{++})^r+\Pi^r(D^{-+})^r
\big] ,\label{D-+} \\
(D^{--})^r &= (D^{-+}_0)^r\Pi^r\big[ (D^{+-})^r + (D^{--})^r \big]\label{D--} .
\end{align}
From the above we infer that $(D^{++})^r\propto \theta(\omega)$ and $(D^{-+})^r\propto \theta(-\omega)$, which leads to:
\begin{equation}
(D^{++})^r = (D_0^{+-})^r\Pi^r (D^{++})^r \label{D++r}.
\end{equation}Similar consideration shows that 
\begin{equation}
(D^{--})^r=(D_0^{-+})^r\Pi^r (D^{--})^r. \label{D--r}
\end{equation}
Since Eqs.~\eqref{D++r} and \eqref{D--r} hold for any self energy $\Pi^r$, we must have $(D^{++})^r=0=(D^{--})^r$. With this, expanding the matrix Keldysh equation $\bm{D^{>,<}}=\bm{D^r}\bm{\Pi^{>,<}}\bm{D^a}$, we find for the greater Green's functions:
\begin{align}
(D^{++})^> &=  (D^{+-})^r    \Pi^>  (D^{-+})^a  ,\label{D++>} \\
(D^{+-})^> &=  (D^{+-})^r \Pi^> (D^{+-})^a , \label{D+->} \\
(D^{-+})^> &=  (D^{-+})^r   \Pi^> (D^{+-})^a  ,\label{D-+>}\\
(D^{--})^> &=  (D^{-+})^r  \Pi^> (D^{+-})^a. \label{D-->}
\end{align}
Together with $[(D^{\pm\mp})^r]^*=(D^{\pm\mp})^a$, we conclude that $(D^{++})^>=(D^{-+})^>=(D^{--})^>=0$.

With the last equations, we proceed to expand \eqref{integralcurrent}. By construction, anti-normal ordering moves the annihilation operator to the left: $|: \phi^-\phi^+:|=\phi^+\phi^-$. We find:
\begin{equation}
\begin{split}
\braket{j(z)} &= \frac{-i\epsilon_0}{2}\int_{\mathbb{R}}\frac{d\omega}{2\pi}\omega \frac{\partial}{\partial z'} \langle\phi^+(z,\omega)\phi^-(z',0)\\&+\phi^+(z',0)\phi^-(z,\omega) \rangle_{z'=z} + \textrm{c.c.}
\end{split} 
\end{equation}
On the $\braket{\phi^+(z',0)\phi^-(z,\omega)}$ term, we perform a variable transformation $\omega\mapsto -\omega$. Using the identity $[\phi^+(z,\omega)]^\dagger=\phi^-(z,-\omega)$, the above becomes:
\begin{equation}
\begin{split}
\braket{j(z)} &= \frac{\epsilon_0}{2}\int_{\mathbb{R}}\frac{d\omega}{\pi}\omega \frac{\partial}{\partial z'}\textrm{Im} \langle\phi^+(z,\omega)\phi^-(z',0)\rangle_{z'=z}\\&+\textrm{c.c.}
\end{split} 
\end{equation}
Noticing that $(D^{+-})^>$ contains only positive frequency: $(D^{+-})^>\propto \theta(\omega)$, the negative half-axis does not contribute to the integral, and we are left with:
\begin{equation}
\begin{split}
\braket{j(z)} &= {\epsilon_0}\int_{0}^{\infty}\frac{d\omega}{\pi}\omega \frac{\partial}{\partial z'}\textrm{Im} \langle\phi^+(z,\omega)\phi^-(z',0)\rangle_{z'=z}.
\end{split} 
\end{equation}
This is already a form suggestive of the greater Green's function. To proceed, consider \begin{equation}
D^>(z,z',\omega)=-\frac{i}{\hbar}\braket{\phi^+(z,\omega)\phi^-(z',0)+\phi^-(z,\omega)\phi^+(z',0)},
\end{equation}where terms with $(D^{++})^>=-\frac{i}{\hbar}\braket{\phi^+\phi^+}$ and $(D^{--})^>=-\frac{i}{\hbar}\braket{\phi^-\phi^-}$ are zero and do not contribute. Observing that $(D^{-+})^>=-\frac{i}{\hbar}\braket{\phi^-\phi^+}=0$, we finally have, for $\omega>0$: $D^>(z,z',\omega)=-\frac{i}{\hbar}\braket{\phi^+(z,\omega)\phi^-(z',0)}$, and hence:
\begin{equation}
\begin{split}
\braket{j(z)} &= {\epsilon_0}\int_{0}^{\infty}\frac{d\omega}{\pi}\omega \frac{\partial}{\partial z'}\textrm{Im}\left[  i\hbar D^>(z,z',\omega)|_{z'=z}\right] \\
&= \epsilon_0  \int_0^{\infty} \frac{d\omega}{\pi} \hbar \omega  \mathrm{Re} \left. \frac{\partial D^>(z,z',\omega)}{\partial z'} \right|_{z'=z}  .
\end{split} 
\end{equation}

\section{Matsubara sum formula for photon self energies due to electrons} \label{Matsubara}
Under the local equilibrium approximation, the polarization diagram, i.e. Eq.~\eqref{photonPi} for the photon self energy, can be calculated exactly. In frequency domain, Eq.~\eqref{photonPi} becomes:
\begin{equation} 
\begin{split}
\Pi^r_{j}(\omega) &= -i\hbar Q^2 \int_{-\infty}^{+\infty}  \frac{d\omega'}{2\pi} \left[ G_{j}^r(\omega') G_{j}^<(\omega'- \omega) \right.\\
& \left. + G_{j}^<(\omega') G^a_{j}(\omega'- \omega) \right].
\end{split}  
\end{equation}
This integral can  be performed by closing a contour using {the} residue theorem. Writing $E=\hbar\omega$, under the wide-band limit for electron bath we find:
\begin{align}
\begin{split}
\Pi^r_j(E) &= Q^2 \Bigg\{ \frac{ i\Gamma_j \bigl[ f(v_j + i \frac{\Gamma_j}{2} + E) - f(v_j + i\frac{\Gamma}{2}) \bigr]}{(E + i\Gamma_j)E}  \\
 & - \sum_{n=0}^\infty  \frac{ik_B T\,\Gamma_j}{(\mu_j -v_j + i\hbar \omega_n)^2+\frac{\Gamma_j^2}{4}}
  \\
 &\quad\times \Big( \frac{1}{\mu_j-v_j + i\hbar \omega_n +E + i \frac{\Gamma_j}{2}} \,  \\
&\quad+  \frac{1}{\mu_j-v_j + i\hbar \omega_n -E - i \frac{\Gamma_j}{2}} \Big) \Bigg\},
\end{split}
\end{align} 
where $f(E)=1/\bigl[\exp(\frac{E-\mu}{k_BT})+1\bigr]$ is the Fermi function, and $\omega_n
=(2n+1)\pi k_B T/\hbar$ is the Matsubara frequency.
\bibliographystyle{apsrev4-1}

\bibliography{RHT_NEGF}

\end{document}